\documentclass[journal,final,twocolumn,twoside]{IEEEtran}
\usepackage[cmex10]{amsmath}
\usepackage{graphicx}
\usepackage{url}
\usepackage{doi}
\usepackage{amssymb}
\usepackage{graphicx}
\usepackage{lineno}
\usepackage{verbatim}
\usepackage{bm}
\usepackage{multirow,textcomp,booktabs}
\usepackage[usenames,dvipsnames]{color}
\usepackage{colortbl}
\usepackage{balance}
\usepackage{times,amsmath,epsfig}
\usepackage{subfigure}
\usepackage[round-mode=places, round-integer-to-decimal, round-precision=1]{siunitx}
\usepackage{mathrsfs}
\usepackage{bm}
\usepackage{cite}
\usepackage{amsmath}

\usepackage{hyperref}
\usepackage{mathrsfs}
\begin{document}
\title{Image Steganography using Gaussian Markov Random Field Model}

\author{Wenkang~Su,~\IEEEmembership{Student~Member,~IEEE,}
        Jiangqun~Ni$^*$,~\IEEEmembership{Member,~IEEE,}
        Yuanfeng~Pan,
        Xianglei~Hu,~\IEEEmembership{Member,~IEEE,}
        Yun-Qing~Shi,~\IEEEmembership{Fellow,~IEEE}
\thanks{$^*$Corresponding author.}}

\maketitle

\begin{abstract}
Recent advances on adaptive steganography show that the performance of image steganographic communication can be improved by incorporating the non-additive models that capture the dependences among adjacent pixels. In this paper, a Gaussian Markov Random Field model (GMRF) with four-element cross neighborhood is proposed to characterize the interactions among local elements of cover images, and the problem of secure image steganography is formulated as the one of minimization of KL-divergence in terms of  a series of low-dimensional clique structures associated with GMRF by taking advantages of the conditional independence of GMRF. The adoption of the proposed GMRF tessellates the cover image into two disjoint subimages, and an alternating iterative optimization scheme is developed to effectively embed the given payload while minimizing the total KL-divergence between cover and stego, i.e., the  statistical detectability. Experimental results demonstrate that the proposed GMRF outperforms the prior arts of model based schemes, e.g., MiPOD, and rivals the state-of-the-art HiLL for practical steganography, where the selection channel knowledges are unavailable to steganalyzers.

\end{abstract}

\IEEEpeerreviewmaketitle

\section{Introduction}
\IEEEPARstart{S}{teganography} is the science and art of covert communication which aims to convey secret messages by slightly modifying the digital media (cover), e.g., images, videos and audios, to create the stego objects.  To ensure the security of the covert communication, the statistical distributions of the cover source should be least possible changed for a given payload. So far, the majority of research works in steganography rely on digital images as covers, either in spatial domain for uncompressed images \cite{WOW, Uniward, HiLL, MVG, Hugo, MVGG, MiPOD} or in DCT domain for JPEG compressed images \cite{UED, UERD, GUED, BET, Micro}.

At present, the mainstream of steganographic schemes for digital images is based on the concept of content adaptivity, i.e., embedding the secret data into the complex regions with rich-texture contents in images. Such adaptive steganographic schemes are generally implemented by first assigning a different embedding cost to each pixel and then embedding the secret data while minimizing the sum of costs of all changed pixels. Although the coding framework, e.g., the syndrome-trellis code (STC) \cite{STC}, is well developed, which allows the steganographers to minimize an additive distortion function under the given payload constraint, most existing distortion functions are said to be heuristically defined, because one can hardly establish a direct connection between distortion and statistical detectability. Currently, several state-of-the-art content-adaptive image steganographic schemes in spatial domain are devised with such heuristically built distortion functions, which include WOW \cite{WOW}, S-UNIWARD \cite{Uniward} and HiLL \cite{HiLL}.

MG \cite{MVG} is the first steganographic scheme which provides a systematic approach for the design of distortion function based on sound mathematical principle. In specific, the distortion is associated with the steganographic Fisher Information \cite{Est-FI}, which is proportional to the Kullback-Leibler (KL) divergence between the statistical distributions of cover and stego images. The authors proposed to model the cover as a sequence of independent quantized multivariate Gaussian random variables with unequal local variances, and the embedding change probabilities for each cover pixel are derived to minimize the total KL-divergence by incorporating with the method of Lagrange multipliers for a given embedding operation (symmetric embedding) and payload. The cost values are then determined with the obtained change probabilities. Although a rather simple variance estimator was employed, the authors showed that the security performance of MG was comparable to the previous state-of-the-art HUGO algorithm \cite{Hugo}.

The authors extended their work of MG later in \cite{MVGG} by replacing the multivariate Gaussian model with the generalized multivariate Gaussian (hence its name MVGG) and incorporating an improved variance estimator. The MVGG was known as model driven in the sense that it allows pentary embedding to hide large payload in complex and texture regions of cover images with a thicker-tail MVGG model. The MVGG showed comparable performance with pentary coded S-UNIWARD and HiLL on SRM \cite{SRM} and maxSRMd2 \cite{maxSRMd2} feature sets.

More recently, following the framework of model-based steganography, an alternative approach \cite{MiPOD} for the design of distortion function was proposed by minimizing the power of optimal detector (MiPOD). Although MiPOD was a continuation of MVGG, it still shed some new insights into steganography design. Firstly, a closed-form expression for the power of the most powerful detector of adaptive embedding with LSB matching is derived based on a multivariate Gaussian cover image model developed in \cite{MVG}. Secondly, the closed-form expression for statistical detectability allows ones to design the so-called “detectability-limited sender” that controls the size of secure payload for a given image to not exceed a target detectability level. Finally, the MiPOD derives the embedding change probabilities (selection channels) for steganography by minimizing the power of optimal detector rather than the KL divergence of the statistical distributions between cover and stego objects as employed in MG and MVGG. Equipped with the improved variance estimator for model parameters, the MiPOD can rival the state-of-the-art steganographic schemes in spatial domain, e.g., S-UNIWARD and HiLL.

In this paper, we go a step further to investigate model that can capture dependencies among spatially adjacent pixels for efficient image steganography with symmetric embedding in spatial domain. In specific, a Gaussian Markov Random Field model (GMRF) with four-element cross neighborhood is proposed to characterize the cover image, which tessellates the cover image set ${\bf{c}}$ into two disjoint subsets (sublattices) ${\bf{c}}_e$ and ${\bf{c}}_o$. For the given GMRF model, each pixel is associated with four two-pixel cliques (two horizontally and two vertically neighboring pixels), which makes up a {\textbf{subtree}} with five pixels. And the involved adjacent pixel pair is modeled as the jointly Gaussian random variables. To embed $L$ bits message in cover $\bf{c}$, we assign $L/2$ bits into sublattices ${\bf{c}}_e$ and ${\bf{c}}_o$ to generate stegos ${\bf{s}}_e$ and ${\bf{s}}_o$, respectively. The Markov Random Field (MRF) modeling is appealing because the subtrees associated with one sublattice, say ${\bf{c}}_e$, are conditionally independent when ${\bf{c}}_o$ is determined, which allows us to formulate the total KL-divergence between ${\bf{c}}$ and ${\bf{s}}$ as the sum of KL-divergence of all subtrees associated with ${\bf{c}}_e$ and ${\bf{c}}_o$ in terms of the ones of 4 neighboring pixel pairs. An alternating iterative strategy among two sublattices is applied to derive the embedding change probabilities (selection channels) and then the distortion costs for each pixel in one sublattice when the embedding payload and the selection channels for another sublattice are given by minimizing the total KL-divergence between cover and stego in the corresponding sublattice. Extensive experiments are carried out to verify the effectiveness of the proposed method (known as GMRF) using steganalyzers with rich model, e.g., SRM and maxSRMd2, on BOSSbase database \cite{bossbase}. The proposed GMRF outperforms its prior art with independent Gaussian model, i.e., MiPOD, in terms of secure embedding capacity for SRM, and can rival the current state-of-the-art one, i.e., HiLL, with symmetric embedding.

The rest of this paper is structured as follows. In Section II, we first introduce the notations and conventions adopted in this paper and review the necessary preliminaries on MRF. Then the cover and stego image models in terms of a pair of adjacent pixels (clique for GMRF with four-element cross neighborhood) based on jointly Gaussian distribution are derived. The GMRF model is then employed in Section III to formulate the problem of image steganography as the one of minimizing the KL-divergence between cover and stego on two disjoint sublattices, which are followed by the experimental results and analysis in Section IV. And finally the conclusion remarks are drawn in Section V.

\section{Notations and Image Modelling}
\subsection{Notations and conventions}
Let us first give the notations and conventions used throughout the paper. The calligraphic fonts will be used solely for sets (occasionally, other symbols may also be used for sets in the interest of ease presentation), random variables will be typeset in capital letters, while their corresponding realizations will be in lower cases. Vectors and matrices will always typeset in bold fonts. Cover and stego images ${\bf{c}} = ({c_{k}})$ and ${\bf{s}} = ({s_{k}})$ are addressed by their one-dimensional index set ${{\cal S}}=\{1,2,\cdots ,K\}$, where ${c_{k}}$ and ${s_{k}}$ $\in\left\{ {0,1,\cdots,G-1}\right\}$ and $G = 256$ for 8-bit grayscale images. We also use the Iverson bracket, $\left[ P \right]$, defined as $\left[ P \right]$=1 when the statement $P$ is true and zero otherwise.

\subsection{Gaussian Markov Random Field (GMRF)}
\begin{figure}[htbp] 
\centering
{\includegraphics[width=0.35\textwidth]{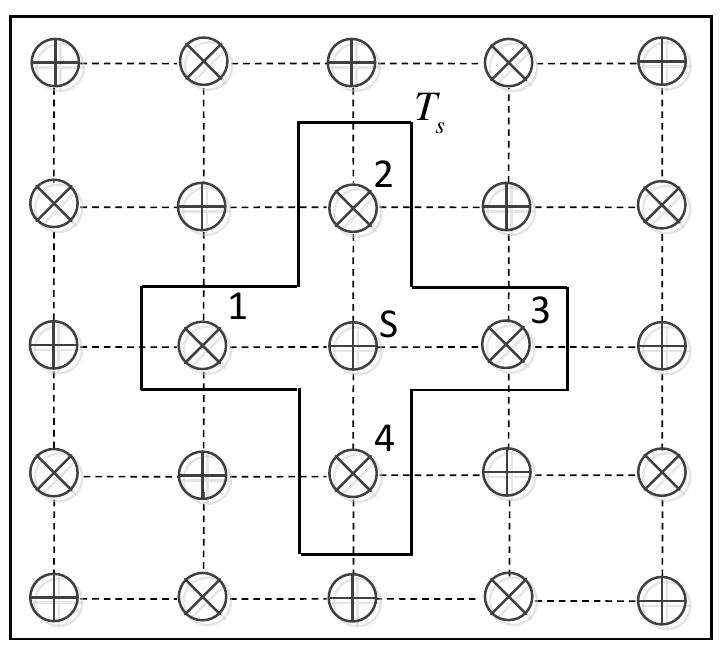}}
\caption{The GMRF model with four-element cross neighborhood.} \label{fig:GMRF_model}
\end{figure}

Let ${\cal S}$ be a set of $n$ discrete sites, i.e., ${{\cal S}} = \{ 1,2, \cdots ,n\}$, and ${\bf{Z}} = ({Z_1},{Z_2}, \cdots ,{Z_n})$ be a family of random variables defined on set ${\cal S}$, the neighborhood system for $\bf{Z}$ is defined as
\begin{equation}\label{eq:GMRF_Neighbor}
{\cal N} = \{ {{\cal N}_k}{\rm{|}}\forall k \in {\cal S}\},
\end{equation}
where ${\cal N}_k$ is the collection of sites neighboring to $k$ for which $k \notin {\cal N}_k$ and $k \in {{{\cal N}}_l} \Leftrightarrow l \in {{{\cal N}}_k}$. In MRF, the ordering of sites are not important, and their relationship is determined by the corresponding neighborhood system ${\cal N}$. A subset ${c} \in {{\cal S}}$ is known as a clique if each pair of different elements from $c$ is neighbor. In this paper, unless stated otherwise, we only consider the four-element cross neighborhood and the associated two-pixel cliques of horizontally or vertically neighboring pixels as shown in Fig. \ref{fig:GMRF_model}.

For the family of random variables $\bf{Z}$, each random variable $Z_k$ takes the value ${z_k} \in {\mathbb{R}}$ with probability $p({Z_k} = {z_k})$, and the joint event $({Z_1} = {z_1}, \cdots ,{Z_n} = {z_n})$ is abbreviated as $\bf{Z}=\bf{z}$ with the probability $p(\bf{Z}=\bf{z})$. $\bf{Z}$ is known as a Markov Random Field (MRF) on ${\cal S}$ with respect to the neighborhood system ${\cal N}$ if and only if (iff) the following two conditions are satisfied \cite{MRF_IMG}:
\begin{equation} \label{eq:GMRF_condition}
\begin{array}{l}
\textbf{Positivity}:  \hspace{0.8cm} p({\bf{Z}}) > 0,  \hspace{0.3cm} \forall {\bf{Z}} \in {{\mathbb{R}}^n}\\
\textbf{Markovianity}:  \hspace{0.2cm}  p({Z_k}|{Z_{{{\cal S}} - \{ k\} }}) = p({Z_k}|{Z_{{{{\cal N}}_k}}})
\end{array}
\end{equation}

The Markovianity says that the probability of a local event at $k$ conditioned on all the remaining events is equivalent to that conditioned on the events at the neighbors of $k$. A random vector ${\bf{Z}} = {({Z_1},{Z_2}, \cdots ,{Z_n})^T} \in {{\mathbb{R}}^n}$ is called a Gaussian MRF (GMRF) w.r.t. a labeled graph ${{\cal G}} = ({{\cal V}},{{\cal E}})$ (${\cal V}$: Vertices; ${\cal E}$: Edges) with mean ${\bm{\mu}}$ and precision matrix ${\bm{\Sigma }}_{}^{ - 1}>0$  , iff it is Gaussian distributed with the form \cite{GMRF}:
\begin{equation} \label{eq:GMRF_Gaussian}
p({\bf{z}})= {((2\pi)^n |{\bm{\Sigma }}|)^{ - {1 \mathord{\left/
 {\vphantom {1 2}} \right.
 \kern-\nulldelimiterspace} 2}}}\exp ( - \displaystyle\displaystyle\frac{1}{2}{({\bf{z}} - {\bm{\mu}} )^T}{\bm{\Sigma }}_{}^{ - 1}({\bf{z}} - {\bm{\mu}} )),
\end{equation}
and ${\bm{\Sigma }}_{k,l}^{ - 1} \ne 0 \Leftrightarrow \left\{ {k,l} \right\} \in {\rm{{\cal E}}}$ for all $k \ne l$.

Once the condition for Positivity is satisfied, which is always true for practical applications, the joint probability $p({\bf{z}})$ is uniquely determined by its local conditional probabilities \cite{MRF_STAT}. In other words, the $p({\bf{z}})$ is characterized with the distributions of all involved cliques associated with the given neighborhood system. We then proceed to develop the cover and stego image model based on the distributions of horizontally or vertically neighboring two-pixel clique in the next two subsections.

\subsection{Cover image model}
As will be discussed later in Section III, a GMRF model with four-element cross neighborhood system is utilized to model the cover images in this paper. For a given uniform scalar quantizer ${Q_\Delta }$ with quantization step ${\Delta}$, the cover is quantized to a zero mean jointly distributed Gaussian ${Q_\Delta }(N({\bf{0}},{\bf{\Sigma }}))$. Without loss of generality, in practice, the mean of the image cover is removed from each pixel, and the ${\Delta}$ is generally set to be 1, in the interest of simplicity. According to \cite{marginal}, for a multivariate Gaussian random variable, its marginal variables are also Gaussian distributed, as a result, both the single pixel and the horizontally or vertically neighboring pixel pair in the cover images, which are associated with the adopted neighborhood system, are assumed to be Gaussian and jointly Gaussian distributed, respectively. Let ${\bf{X}} = \{ {{\bf{X}}^1},{{\bf{X}}^2}, \cdots ,{{\bf{X}}^M}\}$ be the sequence of all involved two-pixel cliques in the cover image, for the $m^{th}$ clique ${{\bf{X}}^m} = {\left[ {X_1^m,X_2^m} \right]^T}$, we have
\begin{equation}\label{eq:Gaussian_1}
{{\bf{X}}^m} \sim N({\bf{0}},{{\bf{\Sigma }}^m}),\hspace{0.3cm} m = 1,2, \cdots ,M,
\end{equation}
that is
\begin{equation}\label{eq:Gaussian_2}
\begin{array}{*{20}{c}}
{{f}(x_1^m,x_2^m) = (({2\pi})^2{{|{{\bf{\Sigma }}^m}|})^{ - {1 \mathord{\left/
 {\vphantom {1 2}} \right.
 \kern-\nulldelimiterspace} 2}}}\exp ( - \displaystyle\displaystyle\frac{1}{2}({\bf{x}}^m)^T({\bf{\Sigma }}^m)^{ - 1}{{\bf{x}}^m})},\\
{{{\bf{\Sigma }}^m} = \left[ {\begin{array}{*{20}{c}}
{\sigma _{1,1}^m}&{\sigma _{1,2}^m}\\
{\sigma _{2,1}^m}&{\sigma _{2,2}^m}
\end{array}} \right]{\rm{  }}\hspace{0.1cm} \text{and} \hspace{0.2cm} {\rm{  }}({\bf{\Sigma }}^m)^{ - 1} = \left[ {\begin{array}{*{20}{c}}
{\gamma _{1,1}^m}&{\gamma _{1,2}^m}\\
{\gamma _{2,1}^m}&{\gamma _{2,2}^m}
\end{array}} \right]},
\end{array}
\end{equation}
where ${{\bf{\Sigma }}^m}$ and $({\bf{\Sigma }}^m)^{ - 1}$ are the \textbf{covariance and precision matrices}, resepctively. The corresponding Probability Mass Function (p.m.f) for the $m^{th}$ clique can be computed under the fine quantization step $\Delta$,
\begin{equation}\label{eq:Gaussian_intergral}
{F_\Delta }(x_1^m,x_2^m) = \int_{x_1^m - \Delta /2}^{x_1^m + \Delta /2} {\int_{x_2^m - \Delta /2}^{x_2^m + \Delta /2} {{f}(x_1^m,x_2^m)dx_2^m} } dx_1^m.
\end{equation}
Let $x_1^m = i\Delta ,x_2^m = j\Delta$, the p.m.f $p_{i,j}^m$ for the $m^{th}$ clique can then be evaluated by incorporating the Mean Value Theorem (MVT):
\begin{equation}\label{eq:MVT}
p_{i,j}^m = {\left. {{F_\Delta }(x_1^m,x_2^m)} \right|_{x_1^m = i\Delta ,x_2^m = j\Delta }} = {\Delta ^2} \cdot f(i'\Delta ,j'\Delta ),
\end{equation}
where $i' \in (i - 0.5,i + 0.5)$ and $j' \in (j - 0.5,j + 0.5)$.

\subsection{Stego image model}
\begin{figure}[htbp] 
\centering
{\includegraphics[width=0.3\textwidth]{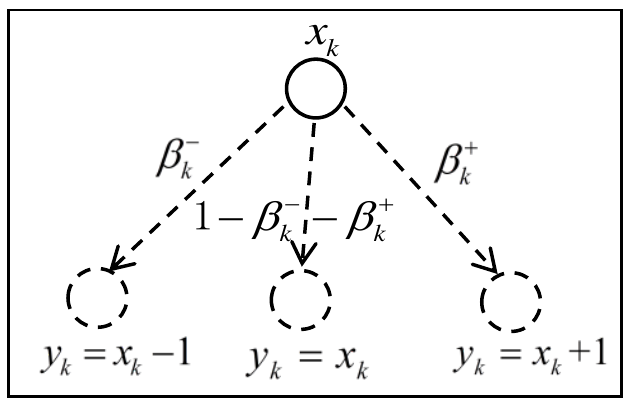}}
\caption{The symmetric ternary embedding model.} \label{fig:stego_model_1}
\end{figure}

Unlike the MG \cite{MVG} and MiPOD \cite{MiPOD} with mutually inde-pendent embedding, we take into account the interactions among embedding changes of neighboring pixels when modifying the cover ${\bf{x}} = ({x_k})$ to stego ${\bf{y}} = ({y_k})$, where ${x_k}$ and ${y_k}$ are the $k^{th}$ cover and stego pixels, respectively.  However, to simplify the problem, the same symmetric ternary embedding model as shown in Fig. \ref{fig:stego_model_1} is adopted throughout the paper, i.e.,
\begin{equation}\label{eq:one_pixel_stego_model}
\begin{array}{l}
{p}({y_k} = {x_k} + 1) = \beta _k^ + \\
{p}({y_k} = {x_k} - 1) = \beta _k^ - \\
{p}({y_k} = {x_k}) = 1 - \beta _k^ +  - \beta _k^ -
\end{array},
\end{equation}
where the change probability $\beta _k^ +  = \beta _k^ -  = {\beta _k}$, ${\beta _k} \in [0,{1 \mathord{\left/{\vphantom {1 3}} \right.\kern-\nulldelimiterspace} 3}]$ is the change probability for the $k^{th}$ pixel of cover $\bf{x}$.

\begin{figure}[htbp] 
\centering
{\includegraphics[width=0.48\textwidth]{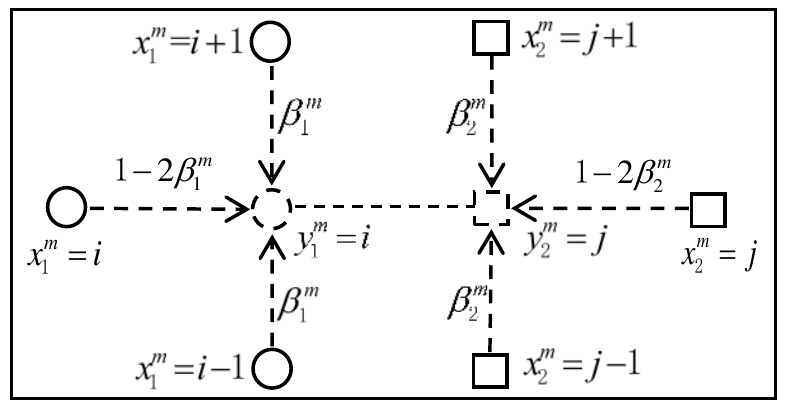}}
\caption{The stego image model for a pair of two-pixel clique with symmetric ternary embedding.} \label{fig:stego_model_2}
\end{figure}

With the symmetric embedding model, the two-pixel cover clique sequence ${\bf{X}} = \{ {{\bf{X}}^m}\} $ with p.m.f ${p^m}(x_1^m,x_2^m)$ is changed to stego clique sequence ${\bf{Y}} = \{ {{\bf{Y}}^m}\}$ with p.m.f ${q^m}(y_1^m,y_2^m)$, where $m = 1,2 \cdots ,M$. Let $\beta _1^m$ and $\beta _2^m$ be the change probabilities for pixels $X_1^m$ and $X_2^m$ in the $m^{th}$ clique, and ${\bf{X}}_{i,j}^m$ and ${\bf{Y}}_{i,j}^m$ denote the $m^{th}$ two-pixel cover and stego cliques which take the value $[i,j]$, respectively. Under ternary embedding with symmetric changes, there are totally 9 embedding change types as shown in Fig. \ref{fig:stego_model_2},
\begin{equation}\label{eq:two_pixels_stego_model}
\begin{array}{l}
{q^m}({\bf{Y}}_{i,j}^m = {\bf{X}}_{i + 1,j}^m + {[ - 1,0]^T}|{\bf{X}}_{i + 1,j}^m) = \beta _1^m(1 - 2\beta _2^m)\\
{q^m}({\bf{Y}}_{i,j}^m = {\bf{X}}_{i,j + 1}^m + {[0, - 1]^T}|{\bf{X}}_{i,j + 1}^m) = (1 - 2\beta _1^m)\beta _2^m\\
{q^m}({\bf{Y}}_{i,j}^m = {\bf{X}}_{i - 1,j}^m + {[1,0]^T}|{\bf{X}}_{i - 1,j}^m) = \beta _1^m(1 - 2\beta _2^m)\\
{q^m}({\bf{Y}}_{i,j}^m = {\bf{X}}_{i,j - 1}^m + {[0,1]^T}|{\bf{X}}_{i,j - 1}^m) = (1 - 2\beta _1^m)\beta _2^m\\
{q^m}({\bf{Y}}_{i,j}^m = {\bf{X}}_{i - 1,j - 1}^m + {[1,1]^T}|{\bf{X}}_{i - 1,j - 1}^m) = \beta _1^m\beta _2^m\\
{q^m}({\bf{Y}}_{i,j}^m = {\bf{X}}_{i + 1,j + 1}^m + {[ - 1, - 1]^T}|{\bf{X}}_{i + 1,j + 1}^m) = \beta _1^m\beta _2^m\\
{q^m}({\bf{Y}}_{i,j}^m = {\bf{X}}_{i + 1,j - 1}^m + {[ - 1,1]^T}|{\bf{X}}_{i + 1,j - 1}^m) = \beta _1^m\beta _2^m\\
{q^m}({\bf{Y}}_{i,j}^m = {\bf{X}}_{i - 1,j + 1}^m + {[1, - 1]^T}|{\bf{X}}_{i - 1,j + 1}^m) = \beta _1^m\beta _2^m\\
{q^m}({\bf{Y}}_{i,j}^m = {\bf{X}}_{i,j}^m + {[0,0]^T}|{\bf{X}}_{i,j}^m) = (1 - 2\beta _1^m)(1 - 2\beta _2^m)
\end{array},
\end{equation}
where $\beta _1^m$ and $\beta _2^m$ $\in \left[ {0,{1 \mathord{\left/ {\vphantom {1 3}} \right. \kern-\nulldelimiterspace} 3}} \right]$. Therefore, the p.m.f for the $m^{th}$ clique in stego image can be written as
\begin{equation}\label{eq:one_pixel_stego_probability}
\begin{array}{l}
q_{i,j}^m = p_{i,j}^m(1 - 2\beta _1^m)(1 - 2\beta _2^m)\\
 + (p_{i - 1,j}^m + p_{i + 1,j}^m)\beta _1^m(1 - 2\beta _2^m)\\
 + (p_{i,j - 1}^m + p_{i,j + 1}^m)(1 - 2\beta _1^m)\beta _2^m\\
 + (p_{i - 1,j - 1}^m + p_{i - 1,j + 1}^m + p_{i + 1,j - 1}^m + p_{i + 1,j + 1}^m)\beta _1^m\beta _2^m
\end{array},
\end{equation}
where $p_{i,j}^m$ and $q_{i,j}^m$ are the probability values for the $m^{th}$ two-pixel cover and stego cliques which take the value $[i,j]$, respectively.

\section{Optimize the Statistical Undetectability by Minimizing the KL-Divergence}
In this Section, we formulate the problem of secure image steganography as the one of minimization of KL-divergence between cover and stego images in terms of a series of low-dimensional clique structures by utilizing the conditional independence of GRMF. The KL-divergence for a two-pixel clique is firstly derived. And the parameter estimation associated with the GMRF is then discussed. Based on these, the total KL-divergence model between cover and stego images is developed and an alternating iterative optimization scheme is proposed to solve the model for the minimization of KL-divergence, i.e., least statistical detectability.

\subsection{The KL-Divergence of a two-pixel clique}
For a two-pixel cover clique ${\bf{X}} = {[{X_1},{X_2}]^T}$, the embedding modifies the two pixels with probabilities ${\boldsymbol{\beta}} = {\left[ {{\beta_1},{\beta_2}} \right]^T}$, changing to the stego clique ${\bf{Y}} = {[{Y_1},{Y_2}]^T}$. Note that we omit the superscript $``m"$ for simplicity. For small change probabilities ${\boldsymbol{\beta}}$, the KL-divergence between cover and stego cliques can be well approximated with its leading quadratic terms (detailed derivations see Appendix A-1):
\begin{equation}\label{eq:two_pixel_KL}
D({\boldsymbol{\beta }}) = {D_{KL}}({F_P}||F_Q^{\boldsymbol{\beta }}) \approx \displaystyle\displaystyle\frac{1}{2}{{\boldsymbol{\beta }}^T} \cdot {\nabla ^2}D({\bf{0}}) \cdot {\boldsymbol{\beta }},
\end{equation}
where ${F_P}$ and $F_Q^{\bm{\beta }}$ are the p.m.f of the two-pixel cover clique and stego clique, respectively, ${\nabla ^2}D({\bf{0}})$ is the second order partial derivative of $D$ w.r.t ${\boldsymbol{\beta}}$ when ${\boldsymbol{\beta}}= \bf{0}$. Similar to the result in \cite{FI-secur}, ${\nabla ^2}D({\bm{\beta}})$ is also proportional to the Fisher information matrix ${{\bf{I}}_2}({\bm {\beta}})$ at ${\bm {\beta}} = {\bf 0}$ (for proof see Appendix A-1), i.e.,
\begin{equation}\label{eq:KL_second_order_partial_derivative}
{\left. {{\nabla ^2}D({\bm{\beta }})} \right|_{{\bm{\beta }} = 0}} = {{{{\left. {{{\bf{I}}_2}(\bm{\beta })} \right|}_{{\bm{\beta }} = 0}}} \mathord{\left/
 {\vphantom {{{{\left. {{{\bf{I}}_2}(\bm{\beta } )} \right|}_{{\bm{\beta }} = 0}}} {\ln 2}}} \right.
 \kern-\nulldelimiterspace} {\ln 2}},
\end{equation}
where ${{\bf{I}}_2}({\bf{0}})$ is the binary steganographic Fisher Information Matrix (FIM)
\begin{equation}\label{eq:FIM}
{{\bf{I}}_2}({\bf{0}}) = \left[ {\begin{array}{*{20}{c}}
{{I_2}{{({\bf{0}})}_{1,1}}}&{{I_2}{{({\bf{0}})}_{1,2}}}\\
{{I_2}{{({\bf{0}})}_{2,1}}}&{{I_2}{{({\bf{0}})}_{2,2}}}
\end{array}} \right],
\end{equation}
where
\begin{equation}\label{eq:FIM_element_1}
\begin{array}{l}
{I_2}{({\bf{0}})_{k,l}} = E\left[ {{{\left. {\left( {\displaystyle\frac{{\partial \ln F_Q^{\bm{\beta }}({\bf{X}})}}{{\partial {\beta _k}}} \cdot \displaystyle\frac{{\partial \ln F_Q^{\bm{\beta }}({\bf{X}})}}{{\partial {\beta _l}}}} \right)} \right|}_{{\bm{\beta }} = {\bf{0}}}}} \right],\\
\hspace{1.6cm} k \in \{ 1,2\}, \hspace{0.2cm} l \in \{ 1,2\} .
\end{array}
\end{equation}
Furthermore, by incorporating \eqref{eq:one_pixel_stego_probability} and \eqref{eq:MVT} for the derivation of \eqref{eq:FIM_element_1} through Taylor expansion, we can derive the closed-form expression for FIM ${{\bf{I}}_2}({\bf{0}})$ (refer to Appendix A-2 for detailed derivations):
\begin{equation}\label{eq:FIM_element_2}
\begin{array}{l}
{{{I}}_2}{({\bf{0}})_{1,1}} = {\left. {\sum\limits_i {\sum\limits_j {\displaystyle\frac{{{{(\Omega _{i,j}^1)}^2}}}{{{p_{i,j}}}}} } } \right|_{{\bm{\beta }} = {\bf{0}}}}\\
{{{I}}_2}{({\bf{0}})_{1,2}} = {{{I}}_2}{({\bf{0}})_{2,1}} = {\left. {\sum\limits_i {\sum\limits_j {\displaystyle\frac{{\Omega _{i,j}^1 \cdot \Omega _{i,j}^2}}{{{p_{i,j}}}}} } } \right|_{{\bm{\beta }} = {\bf{0}}}}\\
{{{I}}_2}{({\bf{0}})_{2,2}} = {\left. {\sum\limits_i {\sum\limits_j {\displaystyle\frac{{{{(\Omega _{i,j}^2)}^2}}}{{{p_{i,j}}}}} } } \right|_{{\bm{\beta }} = {\bf{0}}}}
\end{array},
\end{equation}
where $\Omega _{i,j}^1 = ({p_{i - 1,j}} + {p_{i + 1,j}} - 2{p_{i,j}})$, $\Omega _{i,j}^2 = ({p_{i,j - 1}} + {p_{i,j + 1}} - 2{p_{i,j}})$, and by applying the result in \eqref{eq:MVT}, for fine quantization step $\Delta$, $f(i'\Delta,j'\Delta) \approx f(i\Delta,j\Delta)$, where $i' \in (i - 0.5,i + 0.5)$, $j' \in (j - 0.5,j + 0.5)$, thus we further have
\begin{equation}\label{eq:omiga}
\begin{array}{l}
\Omega _{i,j}^1 \approx {\Delta ^4} \cdot {\left. {\displaystyle\frac{{\partial {}^2{f}({x_1},{x_2})}}{{\partial {{x_1}^2}}}} \right|_{{x_1} = i\Delta ,{x_2} = j\Delta }}\\
\Omega _{i,j}^2 \approx {\Delta ^4} \cdot {\left. {\displaystyle\frac{{\partial {}^2{f}({x_1},{x_2})}}{{\partial {{x_2}^2}}}} \right|_{{x_1} = i\Delta ,{x_2} = j\Delta }}
\end{array},
\end{equation}
where $f(x_1,x_2)$ is the bivariate Gaussian p.d.f as shown in \eqref{eq:Gaussian_2}. It is observed that $\Omega _{i,j}^1$ and $\Omega _{i,j}^2$ are linearly proportional to the second order partial derivatives of ${f}({x_1},{x_2})$ (bivariate Gaussian for two-pixel clique) w.r.t. ${x_1}$ and ${x_2}$ at ${x_1} = i\Delta $ and ${x_2} = j\Delta $, respectively. On the other hand, the second order partial derivatives of ${f}({x_1},{x_2})$ in \eqref{eq:omiga} can be easily obtained according to \eqref{eq:Gaussian_2}, i.e.,
\begin{equation}\label{eq:Guassian_second_order_partial_derivative}
\begin{array}{l}
\displaystyle\frac{{{\partial ^2}{f}({x_1},{x_2})}}{{\partial {x_1}^2}} = {f}({x_1},{x_2}) \cdot {\left[{({\gamma _{1,1}}{x_1} + {\gamma _{1,2}}{x_2})^2} - {\gamma _{1,1}}\right]}\\
\displaystyle\frac{{{\partial ^2}{f}({x_1},{x_2})}}{{\partial {x_2}^2}} = {f}({x_1},{x_2}) \cdot {\left[{({\gamma _{2,1}}{x_1} + {\gamma _{2,2}}{x_2})^2} - {\gamma _{2,2}}\right]}
\end{array},
\end{equation}
where ${\gamma _{1,1}}$, ${\gamma _{1,2}}$, ${\gamma _{2,1}}$ and ${\gamma _{2,2}}$ are the components of the precision matrix ${{\bm{\Sigma }} ^{ - 1}}$ in \eqref{eq:Gaussian_2}. Based on these, substituting \eqref{eq:Guassian_second_order_partial_derivative} into \eqref{eq:omiga} at ${x_1} = i\Delta, {x_2} = j\Delta$, then we can have the FIM ${\bf{I}}_2(\bf{0})$ (${\lim \Delta  \to 0}$)
\begin{equation}\label{eq:FIM_element_3}
\begin{array}{l}
{I_2}{{\rm{(}}{\bf{0}}{\rm{)}}_{{1,1}}} = \sum\limits_i {\sum\limits_j {\displaystyle\frac{{{{{\rm{(}}\Omega _{i,j}^1{\rm{)}}}^2}}}{{{p_{i,j}}}}} } \\
\approx { {\sum\limits_i {\sum\limits_j {\displaystyle\frac{{{{\left( {{\Delta ^4}\cdot f{\rm{(}}i\Delta ,j\Delta {\rm{)}} \cdot \left[ {{{{\rm{(}}{\gamma _{1,1}}i\Delta  + {\gamma _{1,2}}j\Delta {\rm{)}}}^2} - {\gamma _{1,1}}} \right]} \right)}^2}}}{{{\Delta ^2} \cdot f{\rm{(}}i\Delta ,j\Delta {\rm{)}}}}} } } }\\
{\mathop \approx} {\Delta ^4} \cdot \displaystyle {\iint_{{{\mathbb{R}}^2}}{f}({x_1},{x_2})\cdot{{\left[ {{{({\gamma _{1,1}}{x_1} + {\gamma _{1,2}}{x_2})}^2}} - {\gamma _{1,1}}\right]}^2}} \mathrm{d}x_1\mathrm{d}x_2,\\
\end{array}
\end{equation}
\begin{equation}\label{eq:FIM_element_4}
\begin{array}{l}
{{{I}}_2}{({\bf{0}})_{2,2}} = \sum\limits_i {\sum\limits_j {\displaystyle\frac{{{{(\Omega _{i,j}^2)}^2}}}{{{p_{i,j}}}}} }\\
{\mathop \approx} {\Delta ^4} \cdot \displaystyle {\iint_{{{\mathbb{R}}^2}}{f}({x_1},{x_2})\cdot{{\left[ {{{({\gamma _{2,1}}{x_1} + {\gamma _{2,2}}{x_2})}^2}} - {\gamma _{2,2}}\right]}^2}} \mathrm{d}x_1\mathrm{d}x_2,\\
\end{array}
\end{equation}
\begin{equation}\label{eq:FIM_element_5}
\begin{array}{l}
{{{I}}_2}{({\bf{0}})_{1,2}} = {{{I}}_2}{{({\bf{0}})}_{2,1}} = \sum\limits_i {\sum\limits_j {\displaystyle\frac{{\Omega _{i,j}^1 \cdot \Omega _{i,j}^2}}{{{p_{i,j}}}}} } \\
{\mathop \approx} {\Delta ^4} \cdot \displaystyle {\iint_{{{\mathbb{R}}^2}}{f}({x_1},{x_2})\cdot{\left[ {{{({\gamma _{1,1}}{x_1} + {\gamma _{1,2}}{x_2})}^2}} - {\gamma _{11}}\right]}} \\
\hspace{3cm} \cdot {{\left[ {{{({\gamma _{2,1}}{x_1} + {\gamma _{2,2}}{x_2})}^2}} - {\gamma _{2,2}} \right]}}\mathrm{d}x_1\mathrm{d}x_2.
\end{array}
\end{equation}

It is readily seen that the elements of FIM in \eqref{eq:FIM} can be represented in terms of ${2^{th}}$ order and ${4^{th}}$ order moments of bivariate Gaussian $\bf{X}$, which, according to Isserlis' theorem \cite{higher-order-moments}, can be represented as
\begin{equation}\label{eq:Higher moments}
\begin{array}{l}
{\rm{ }}E[X_1^2] = {\sigma _{1,1}}{\rm{ ,}}E[{X_1}{X_2}] = {\sigma _{1,2}},\\
E[X_1^4] = 3{({\sigma _{1,1}})^2},{\rm{ }}E[X_1^3{X_2}] = 3{\sigma _{1,1}}{\sigma _{1,2}},\\
E[X_1^2X_2^2] = {\sigma _{1,1}}{\sigma _{2,2}} + 2{({\sigma _{1,2}})^2},
\end{array}
\end{equation}
where ${\sigma _{1,2}}$ is the covariance of $X_1$ and $X_2$, ${\sigma _{1,1}}$ and ${\sigma _{2,2}}$ are the variance of $X_1$ and $X_2$, respectively. Therefore, we finally obtain the FIM ${{\bf {I}}_2}({\bf 0})$ as follows:
\begin{equation}\label{eq:FIM_element_4}
\begin{array}{l}
{I_2}{{\rm{(}}{\bf{0}}{\rm{)}}_{1,1}} \approx \displaystyle\frac{{2{\Delta ^4}}}{{{{{\rm{(}}{\sigma _{1,1}}{\rm{)}}}^2} \cdot {{\left( {1 - {{{\rm{(}}{\rho _{1,{\rm{2}}}}{\rm{)}}}^2}} \right)}^2}}}\\
{I_2}{{\rm{(}}{\bf{0}}{\rm{)}}_{1,2}} = {I_2}{{\rm{(}}{\bf{0}}{\rm{)}}_{2,1}} \approx \displaystyle\frac{{2{\Delta ^4} \cdot {{{\rm{(}}{\rho _{1,{\rm{2}}}}{\rm{)}}}^2}}}{{{\sigma _{1,1}} \cdot {\sigma _{2,2}} \cdot {{\left( {1 - {{{\rm{(}}{\rho _{1,{\rm{2}}}}{\rm{)}}}^2}} \right)}^2}}}\\
{I_2}{{\rm{(}}{\bf{0}}{\rm{)}}_{2,2}} \approx \displaystyle\frac{{2{\Delta ^4}}}{{{{{\rm{(}}{\sigma _{2,2}}{\rm{)}}}^2} \cdot {{\left( {1 - {{{\rm{(}}{\rho _{1,{\rm{2}}}}{\rm{)}}}^2}} \right)}^2}}}
\end{array},
\end{equation}
where ${\rho _{1,2}} = {{{\sigma _{1,2}}} \mathord{\left/
 {\vphantom {{{\sigma _{1,2}}} {\sqrt {{\sigma _{1,1}}{\sigma _{2,2}}} }}} \right.
 \kern-\nulldelimiterspace} {\sqrt {{\sigma _{1,1}} \cdot {\sigma _{2,2}}} }}$ is the correlation coefficient of $X_1$ and $X_2$.

Thus far, we can rewrite the KL-divergence between a pair of two-pixel cover and stego cliques by taking advantage of \eqref{eq:two_pixel_KL}, \eqref{eq:KL_second_order_partial_derivative}, \eqref{eq:FIM} and \eqref{eq:FIM_element_4},
\begin{equation}\label{eq:two_pixel_KL_2}
{D_{KL}}({F_P}||F_Q^{\bm{\beta }}) = \displaystyle\frac{{({I_2}{{({\bf{0}})}_{1,1}}{\beta _1}^2 + 2{I_2}{{({\bf{0}})}_{1,2}}{\beta _1}{\beta _2} + {I_2}{{({\bf{0}})}_{2,2}}{\beta _2}^2)}}{{2\ln 2}}.
\end{equation}

It is observed that the KL-divergence of a two-pixel clique is relevant to the correlation coefficient, variances and change probabilities of pixels in the clique, among which, the correlation coefficients for each neighboring pixel pair in the involved cliques are the predominant parameters to be determined for the incorporated Gaussian MRF model. In general, the performance of the proposed MRF based steganographic scheme is heavily dependent on the estimation accuracy of the model parameters. To this end, we follow in spirit the scheme in MiPOD \cite{MiPOD} to estimate the pixel variance and the correlation coefficient of neighboring pixels in a clique, which will be illustrated later in next subsection.

\subsection{The estimation of pixel variance and correlation coefficient of neighboring pixels}
In \cite{MiPOD}, Sedighi \emph{et al}. proposed an elegant pixel variance estimator to build the underlying multivariate independent Gaussian model, which leads to notable performance improvement over its predecessor MG \cite{MVG}. The variance estimator consists of two steps, that is: 1) suppress the image content using a denoising filter to obtain the residual image; 2) fit a local parametric model to the neighbors of each residual to obtain its variance estimation.

In light of its effectiveness, the same strategy is also generalized to develop the covariance estimator for the involved Gaussian MRF model. For the 8-bit grayscale cover image with original pixel values ${\bf{c}} = ({c_1},{c_2}, \cdots ,{c_n})$, we obtain the residual ${\bf{r}}$ of the cover ${\bf{c}}$ using a two-dimensional wiener filter $F$: ${\bf{r}} = {\bf{c}} - F({\bf{c}})$. Then we utilize a local parametric model to estimate the pixel variance by blockwise Maximum Likelihood Estimation (MLE) \cite{MLE}. For the ${n^{th}}$ residual ${{{r}}_n}$ in cover ${\bf{c}}$, we fit the local parametric model \cite{local_para} to the $p \times p$ neighbors of ${{{r}}_n}$ and model the residual expectation within the associated $p \times p$ block as follows:
\begin{equation}\label{eq:residual_estimate_1}
{{\bf{r}}_n} = {\bf{G}}{{\bf{a}}_n} + {{\bm \xi} _n},
\end{equation}
where ${{\bf{r}}_n}$ denotes the residual values inside the $p \times p$ block surrounding the ${n^{th}}$ residual ${{{r}}_n}$, which is a column vector of size ${p^2} \times 1$, ${\bf{G}}$ is the matrix of size ${p^2} \times q$ which defines the adopted parametric model, ${{\bf{a}}_n}$ is the parameter vector of size $q \times 1$, and ${{\bm \xi} _n}$ is the noise vector of the model with the variance and covariance to be estimated. Under the assumption of Gaussian noise, the estimation of the residual ${{\bf{r}}_n}$ is
\begin{equation}\label{eq:residual_estimate_2}
{{\bf{\hat r}}_n} = {\bf{G}}{({{\bf{G}}^T}{\bf{G}})^{ - 1}}{{\bf{G}}^T}{{\bf{r}}_n}.
\end{equation}
In general, we could assume the residuals within the $p \times p$ block centered on ${r_n}$ exhibit the similar statistical characteristics, i.e., they could be regarded as the multiple samples of ${r_n}$. Therefore, the covariance of neighboring pixels ${{{c}}_m}$ and ${{{c}}_n}$ in a clique can be well estimated as
\begin{equation}\label{eq:covariance}
{\sigma _{m,n}} = {{{{({{\bf{r}}_m} - {{{\bf{\hat r}}}_m})}^T }({{\bf{r}}_n} - {{{\bf{\hat r}}}_n})} \mathord{\left/
 {\vphantom {{{{({{\bf{r}}_m} - {{{\bf{\hat r}}}_m})}^T }({{\bf{r}}_n} - {{{\bf{\hat r}}}_n})} {({p^2}}}} \right.
 \kern-\nulldelimiterspace} {({p^2}}} - q).
\end{equation}
Similarly, we have the variance estimation for pixel ${{{c}}_n}$, i.e.,
\begin{equation}\label{eq:variance}
{\sigma _{n,n}} = {{{{({{\bf{r}}_n} - {{{\bf{\hat r}}}_n})}^T }({{\bf{r}}_n} - {{{\bf{\hat r}}}_n})} \mathord{\left/
 {\vphantom {{{{({{\bf{r}}_n} - {{{\bf{\hat r}}}_n})}^T }({{\bf{r}}_n} - {{{\bf{\hat r}}}_n})} {({p^2}}}} \right.
 \kern-\nulldelimiterspace} {({p^2}}} - q).
\end{equation}
Given the variance and covariance estimations for pixels ${{{c}}_m}$ and ${{{c}}_n}$, we can readily obtain the correlation coefficient between them, i.e.,
\begin{equation}\label{eq:correlation coefficient}
\rho  = {{{{{\sigma }}_{m,n}}} \mathord{\left/
 {\vphantom {{{{{\sigma }}_{m,n}}} {\sqrt {{{{\sigma }}_{m,m}} \cdot {{{\sigma }}_{n,n}}} }}} \right.
 \kern-\nulldelimiterspace} {\sqrt {{{{\sigma }}_{m,m}} \cdot {{{\sigma }}_{n,n}}} }}.
\end{equation}

Considering the numerical stability and computational efficiency, we set adequate bounds for $\rho$ and ${\sigma _{n,n}}$, i.e.,
$\rho  = min (\rho , 0.99*sign(\rho ))$ and ${\sigma_{n,n}} = max (0.01,{\sigma _{n,n}})$.

\subsection{Minimize the total KL-divergence between cover and stego by incorporating the Gaussian MRF}
With the formulation of KL-divergence of a two-pixel clique in \eqref{eq:two_pixel_KL_2}, we then further derive the total KL-divergence between cover and stego by incorporating the given Gaussian MRF with four-element cross neighborhood as shown in Fig. \ref{fig:GMRF_model}, where the image of size $M \times N$ is decomposed into two interleaved subimages (sublattices) ${\cal A}$ (``$+$'') and  ${\cal B}$ (``$\times$''), i.e.,
\begin{equation}\label{eq:interleaved subimages}
\begin{array}{l}
{{\cal A}} = \{ k, |\bmod (k,2) = 1\} \\
{{\cal B}} = \{ k, |\bmod (k,2) = 0\}
\end{array},
\end{equation}
where $1 \le k \le M \times N$.

For any pixel ${X_{s}}$ in subimage ${\cal A}$ (the same for ${\cal B}$), there are four cliques associated with it, i.e., ${{\bf{X}}_{{s},{t}}}=[{X_s},{X_t}]^T$, $t = 1, \cdots ,4$, which constitute a 4-ary clique tree denoted by ${T_s}=\{X_s,X_1,X_2,X_3,X_4\}$, as shown in Fig. \ref{fig:GMRF_model}. With given ${X_{s}}$ (or ${\beta _{s}}$) in ${\cal A}$, the other four pixels (${X_{t}}$) in cliques ${{\bf{X}}_{{s},{t}}}$ ($t = 1, \cdots ,4$) of ${T_s}$ are conditionally independent under the proposed Gaussian MRF model  with four-element cross neighborhood. We then obtain the KL-divergence for clique tree ${T_s}$ (refer to Appendix B for details)
\begin{equation}\label{eq:tree_KL-divergence}
D_{KL}^{{T_s}} = \displaystyle\frac{{\sum\limits_{c \in {{\cal C}}} {([{\theta _c}]{\bm{\beta }}_c^T{{\bf{I}}_{2,c}}({\bf{0}}){{\bm{\beta }}_c}} ) - ((\sum\limits_{c \in {{\cal C}}} {[{\theta _c}]} ) - 1){I_{1,{X_s}}}(0){{\beta _{X_s}}^2}}}{{2\ln 2}},
\end{equation}
where ${{\cal C}} = \left\{ {{\rm{\{ }}{X_s},{X_t}{\rm{\} }},|t = 1, \cdots 4} \right\}$, ${{\bm{\beta }}_{c}} = {[{\beta _{X_s}},{\beta _{X_t}}]^T}$, ${{\bf{I}}_{2,{c}}}({\bf{0}})$ is binary FIM for clique ${c}$, $I_{1,{{X_s}}}(0)$ is the Fisher Information for pixel $X_{s}$, and $[{\theta _{c}}] = 1$ when clique ${c}$ is included in clique tree $T_s$ and zero otherwise.

Note that, with the underlying GMRF, for given change probabilities for pixels in ${\cal B}$, the pixels in ${\cal A}$ and their associated 4-ary clique trees are mutually independent. And so are the pixels in ${\cal B}$. Let the total payload be $L$ bits, we assign half of the payload to ${\cal A}$ and ${\cal B}$, respectively. Without loss of generality, we take the embedding in ${\cal A}$ for example, the allowable payload that can be embedded into subimage ${\cal A}$ is the sum of entropies of $\left\{ {\beta _{X_s}^{{\cal A}},\beta _{X_s}^{{\cal A}},1 - 2\beta _{X_s}^{{\cal A}}} \right\}$, i.e.,
\begin{equation}\label{eq:payload}
\sum\limits_{X_s} {h(\beta _{X_s}^{{\cal A}})}  = {L \mathord{\left/
 {\vphantom {m 2}} \right.
 \kern-\nulldelimiterspace} 2},
\end{equation}
where $\beta _{X_s}^{{\cal A}}$ is the change probability for the ${s^{th}}$ pixel in ${\cal A}$, and $h(x) =  - 2x\log x - (1 - 2x)\log (1 - 2x)$ is expressed in bits. The minimum distortion embedding for ${\cal A}$ in terms of $\beta _{X_s}^{{\cal A}}$ is then formulated as the minimization of total KL-divergence between cover and stego of subimage ${\cal A}$ subject to the payload constraint \eqref{eq:payload}, which can be solved using the method of Lagrange multipliers,
\begin{equation}\label{eq:objective_func_A}
{\bm{\beta} ^{{\cal A}}}{\rm{ = }}\mathop {{\rm{arg min}}}\limits_{\beta _{X_s}^{{\cal A}}} {\rm{ }}\left\{ {\sum\limits_{T_s} {D_{KL}^{{{\cal A}},T_s}}  - \lambda \left[ {\sum\limits_{X_s} {h(\beta _{X_s}^{{\cal A}}) - {L \mathord{\left/ {\vphantom {m 2}} \right. \kern-\nulldelimiterspace} 2}} } \right]} \right\},
\end{equation}
where $D_{KL}^{{{\cal A}},T_s}$ is the KL-divergence expressed in \eqref{eq:tree_KL-divergence} for the ${s^{th}}$ clique tree in ${\cal A}$. Differentiating the objective w.r.t. $\beta _{X_s}^{{\cal A}}$ gives (refer to Appendix C):
\begin{equation}\label{eq:objective_func_Lagrange}
{\Gamma _{X_s}}\beta _{X_s}^{{\cal A}} + {\Lambda _{X_s}} - 2\lambda \ln \displaystyle\frac{{1 - 2\beta _{X_s}^{{\cal A}}}}{{\beta _{X_s}^{{\cal A}}}} = 0,
\end{equation}
where ${\Gamma _{X_s}} = \sum\limits_{{c} \in {{\cal C}}} {[\theta _{c}^{T_s}]{\bf{I}}_{2,{c}}^{T_s}{{({\bf{0}})}_{2,2}} - (} (\sum\limits_{{c} \in {{\cal C}}} {[\theta _{c}^{T_s}]} ) - 1)I_{1,{X_s}}^{T_s}(0)$ and ${\Lambda _{X_s}} = \sum\limits_{{c} \in {{\cal C}}} {[\theta _{c}^{T_s}]{\bf{I}}_{2,{c}}^{T_s}{{({\bf{0}})}_{1,2}}\beta _{X_{s\_c}}^{{\cal B}}}$, $\beta _{X_{s\_c}}^{{\cal B}}$ is the change probability for pixel ${X_{s\_c}}$ corresponding to the clique $c = \{ {X_s},{X_{s\_c}}\} {\rm{ }}$ of the 4-ary clique tree $T_s$, which is located in subimage $\cal B$ and is fixed in the optimization process, ${\bf{I}}_{2,{c}}^{T_s}({\bf{0}})$ and $I_{1,{X_s}}^{T_s}(0)$ are the binary FIM for clique ${c}$ and FI associated with the ${s^{th}}$ pixel in ${\cal A}$, respectively. The algorithm in \cite{binary-search} is then adopted to obtain $\bm{\beta}^{{\cal A}}$.

Similarly, with the solved $\bm{\beta}^{{\cal A}}$, we have the optimal embedding for subimage ${\cal B}$ to obtain $\bm{\beta}^{{\cal B}}$
\begin{equation}\label{eq:objective_func_B}
{\bm{\beta} ^{{\cal B}}}{\rm{ = }}\mathop {{\rm{arg min}}}\limits_{\beta _{X_s}^{{\cal B}}} \left\{ {\sum\limits_{T_s} {D_{KL}^{{{\cal B}},T_s}}  - \lambda \left[ {\sum\limits_{X_s} {h(\beta _{X_s}^{{\cal B}}) - {L \mathord{\left/
 {\vphantom {L 2}} \right.
 \kern-\nulldelimiterspace} 2}} } \right]} \right\}.
\end{equation}

\begin{figure}[htbp] 
\centering
{\includegraphics[width=0.4\textwidth]{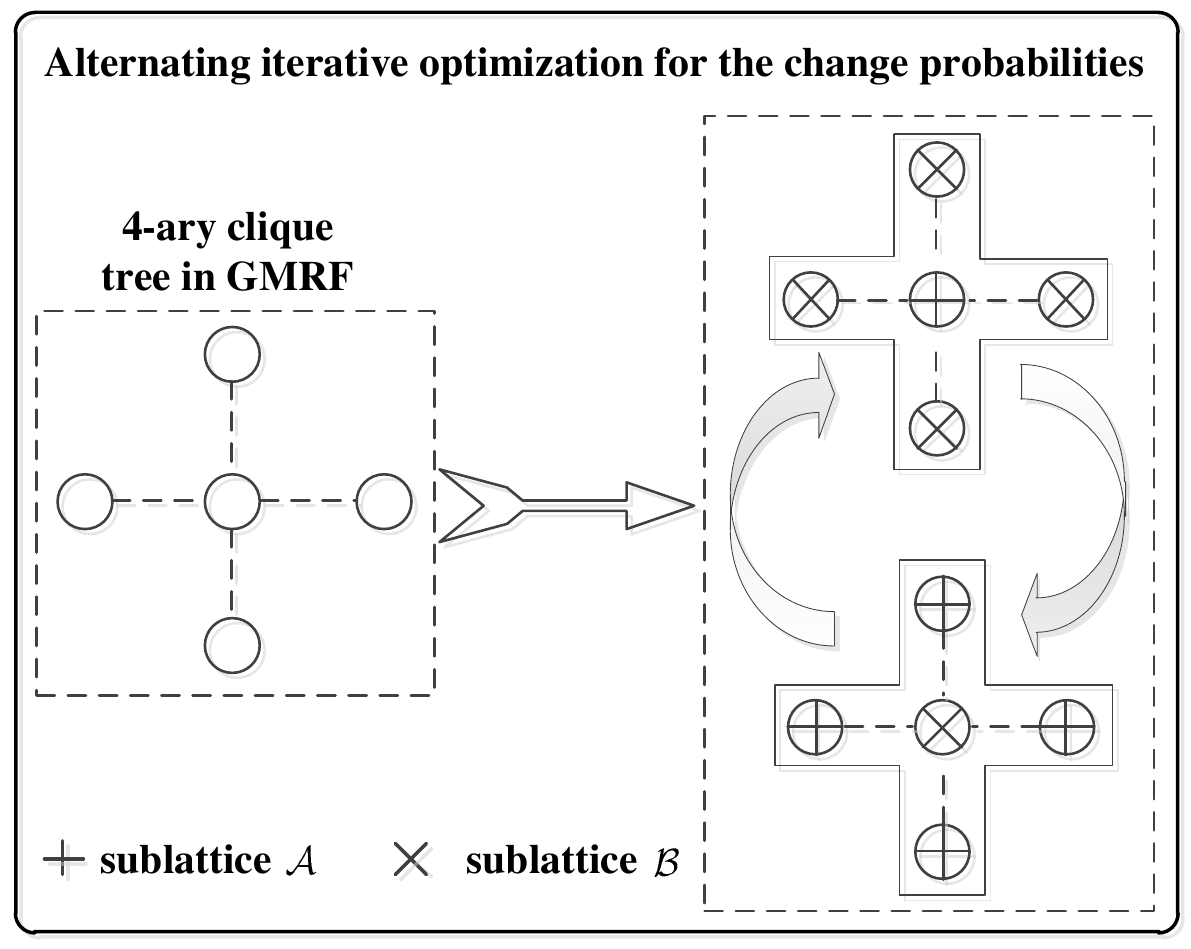}}
\caption{The alternating iterative optimization scheme for optimal change probabilities by incorporating the proposed GMRF model.} \label{fig:Iterative_optimization_2}
\end{figure}

Therefore, an alternating iterative optimization scheme is developed to embed $L$ bits in cover image while minimizing the total KL-divergence by incorporating the proposed GMRF model as shown in Fig. \ref{fig:Iterative_optimization_2}. In specific, randomly initialize the $\bm{\beta}^{{\cal B}}$ in subimage ${\cal B}$ and keep them unchanged, solve \eqref{eq:objective_func_A} for $\bm{\beta}^{{\cal A}}$ in ${\cal A}$. Update the change probabilities for pixels in ${\cal A}$ with the newly obtained $\bm{\beta}^{{\cal A}}$ and then keep them unchanged, solve \eqref{eq:objective_func_B} for $\bm{\beta}^{{\cal B}}$ in ${\cal B}$. The process continues until it converges to obtain the optimal $\bm{\beta}^{{\cal A}}$ and $\bm{\beta}^{{\cal B}}$ for embedding payload of $L$ bits into cover image with minimum distortion. It is noted that: 1) the update of $\bm{\beta}^{{\cal A}}$ is affected somehow by the neighboring $\bm{\beta}^{{\cal B}}$ through the underlying GMRF model; 2) the problems specified in \eqref{eq:objective_func_A} and \eqref{eq:objective_func_B} are the ones of convex optimization, which are bound to converge to the globally optimal solutions. The readers are advised to refer to the pseudo-code (\textbf{Algorithm 1}) with practical consideration to better understand the process.

\begin{figure} [h]
\centering
{
\subfigure[]{\label{fig:clique_allocation_1}\includegraphics[width=0.2\textwidth]{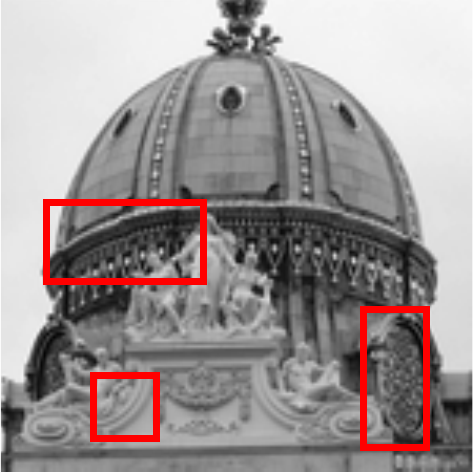}}
}
{
\subfigure[]{\label{fig:clique_allocation_2}\includegraphics[width=0.2\textwidth]{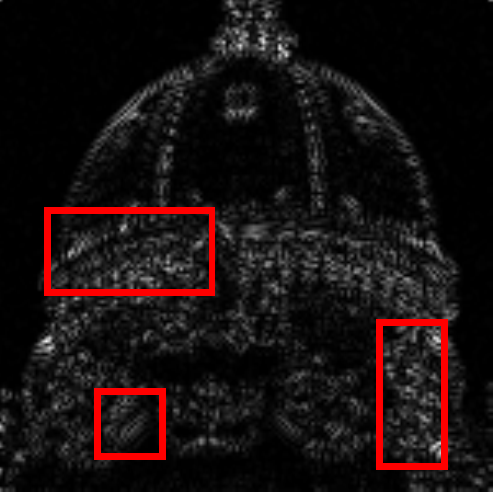}}
}
{
\subfigure[]{\label{fig:clique_allocation_3}\includegraphics[width=0.2\textwidth]{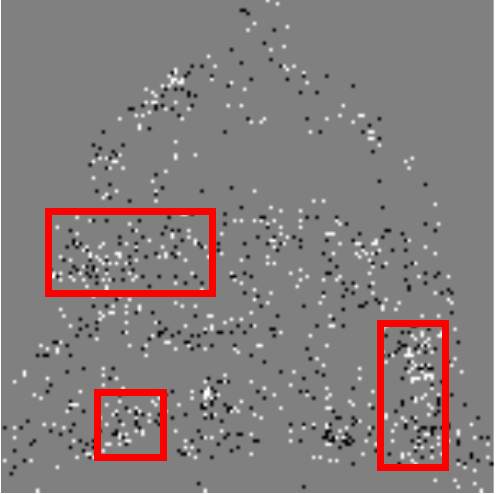}}
}
{
\subfigure[]{\label{fig:clique_allocation_4}\includegraphics[width=0.2\textwidth]{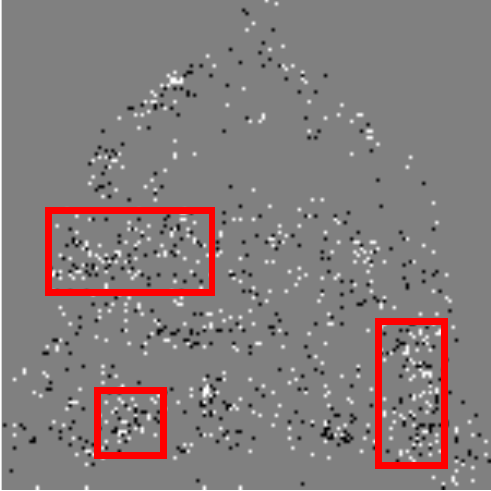}}
}
{
\subfigure[]{\label{fig:clique_allocation_5}\includegraphics[width=0.2\textwidth]{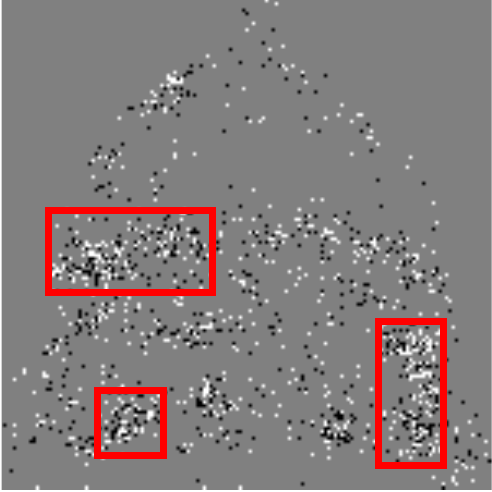}}
}
{
\subfigure[]{\label{fig:clique_allocation_6}\includegraphics[width=0.2\textwidth]{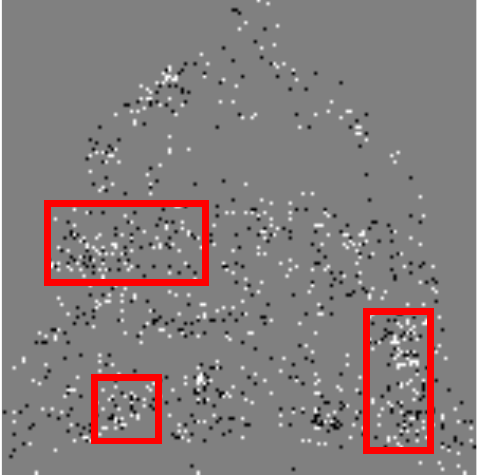}}
}
\caption{Illustration of the modified locations (from (c) to (f)) in a portion of image (1013.pgm) from BOSSbase \cite{bossbase} at payload 0.1 bpp with simulation embedding (the rand seeds of the simulator are fixed). The black, white and gray points represent change `-1', change `+1' and no change, respectively. (a) A portion of cover image `1013.pgm'. (b) The texture regions in the image. (c) The modified locations with MiPOD. (d) The modified locations with the proposed GMRF. (e) The modified locations with the proposed GMRF using the dynamical allocation scheme. (f) The modified locations with the proposed GMRF when all the cliques are excluded in the clique trees.}\label{fig:clique_allocation}
\end{figure}

We then proceed to consider the practical issues in implementing the GMRF based image steganography. Recall that the interactions between neighboring pixels are explored by taking advantages of the proposed GMRF with four-element cross neighborhood, which doesn't go well enough for embedding in rich texture regions of cover images. Fig. \ref{fig:clique_allocation} shows a portion of image (1013.pgm) from BOSSbase \cite{bossbase}, where the red boxes correspond to the texture regions and the black and white points inside images represent the '$-$1' and '$+$1' embeddings, respectively. The paradigm of  adaptive image steganography encourages more data to be embedded in the texture regions of images by assigning low costs to changes to regions which exhibit less correlations among neighboring pixels. The red boxes in  Fig. \ref{fig:clique_allocation} (c) shows the embedding distribution using MiPOD \cite{MiPOD} with a multivariate independent Gaussian model, which is appropriate to characterize the rich texture regions of images, while  Fig. \ref{fig:clique_allocation} (d) shows slightly sparser distributions in the same regions with the proposed GMRF, where all the four two-pixel cliques are included in the clique tree. This is because, for the two less correlated pixels in a clique, say, $X_1$ (low cost) and $X_2$ (high cost), the embedding cost for $X_1$ would increase during the alternating iterative optimization process, which may lead to less embeddings in the texture regions. Therefore, if two horizontally/vertically neighboring pixels are less correlated to some extents, the relevant clique should be disconnected from the GMRF model. Note that the correlation between two neighboring pixels is in connection with their variances, the involved two pixels are less correlated as long as one or two of them have relatively large variances. On the other hand, the ultimate goal for GMRF based optimization is to find the optimal change probability $\beta$ of each pixel $x$ in the cover image for minimum distortion embedding, while $\beta$ is directly proportional to pixel's variance, thus the change probability $\beta$ can be used as an effective measure to dynamically allocate the relevant cliques in the alternating iterative optimization process, i.e., let ${{\bf{X}}_{c}} = {[{X_s},{X_t}]^T}$ be the clique ${c}=\{{X_s},{X_t}\}$ ($t=1,..,4$) associated with the ${s^{th}}$ clique tree, we have
\begin{equation}\label{eq:clique_allocation}
\theta _{c}^{T_s} = \left\{ \begin{array}{l}
1, \hspace{0.3cm}{\beta _{X_s}} \ge {\beta _T} \hspace{0.1cm}{\text{and}}\hspace{0.1cm} {\beta _{X_t}} \ge {\beta _T}\\
0, \hspace{0.3cm}{\text{otherwise}}
\end{array} \right.,
\end{equation}
where $\theta _{c}^{T_s}$ is the statement to determine if clique ${{\bf{X}}_{c}}$ is included in the clique tree $T_s$, ${\beta _T}$ is the threshold, ${\beta _{X_s}}$ and ${\beta _{X_t}}$ are the change probabilities for the two pixels in clique ${{\bf{X}}_{c}}$, respectively. The introduction of \eqref{eq:clique_allocation} can then be regarded as the dynamical configuration of the initial neighborhood system ${\cal N}$ for GMRF based embedding with practical consideration. Fig. \ref{fig:clique_allocation} (e) and (f) show the embedding distributions of GMRF in texture regions when the dynamical allocation scheme is applied and all the cliques are excluded in the clique trees, respectively. It is observed that the GMRF has denser embedding densities than MiPOD in red blocks, while shows similar densities with MiPOD in highly-textured regions. Note that the red blocks in Fig. \ref{fig:clique_allocation} may consists of highly and medium textured regions, and the performance gains of the proposed GMRF over independent Gaussian model based schemes are mainly from the medium and less textured regions in images.

As a summary, we finally give the pseudo-code (\textbf{Algorithm 1}) for GMRF based embedding with practical consideration.
\begin{table}[h]
\renewcommand\arraystretch{1.5}
\begin{tabular}{c}
\toprule[1pt]
\parbox{.45\textwidth}{{\textbf{Algorithm 1}} \hspace{0.3cm} {{Pseudo-code for GMRF based embedding}}.}  \\
\midrule[0.8pt]
\parbox{.45\textwidth}{{\bf {Require:}} the ${{\bf{\sigma }}_{n,n}}$ of a pixel, the $\rho$ associated with a two-pixel clique,} \\
\parbox{.45\textwidth}{\hspace{0cm}embedding payload $L$ bits, mutually disjoint cover subimages: ${\cal A} \cup {\cal B}$.} \\
\parbox{.45\textwidth}{\hspace{0.2cm}{{1.}} Initialize the change probabilities ${\bm{\beta} ^{{\cal B}}}$ in the range $[0,0.001]$.}\\
\parbox{.45\textwidth}{\hspace{0.2cm}{{2.}} \textbf{for} $k = 1$ to $4$ \textbf{do} \hspace{0.2cm}(\emph{GMRF with dynamical clique allocation})} \\
\parbox{.45\textwidth}{\hspace{0.2cm}{{3.}}\hspace{0.4cm}a) Keep the ${\bm{\beta} ^{{\cal B}}}$ unchanged, optimize the ${\bm{\beta} ^{{\cal A}}}$ using \eqref{eq:objective_func_A}, i.e., }\\
\parbox{.45\textwidth}{\hspace{0.8cm}{ } \hspace{0.2cm}${\bm{\beta}^{{\cal A}}}{\rm{ = }}\mathop {{\rm{arg min}}}\limits_{\beta _{X_s}^{{\cal A}}} \left\{ {\sum\limits_{T_s} {D_{KL}^{{{\cal A}},T_s}}  - \lambda \left[ {\sum\limits_{X_s} {h(\beta _{X_s}^{{\cal A}}) - {L \mathord{\left/ {\vphantom {m 2}} \right. \kern-\nulldelimiterspace} 2}} } \right]} \right\}$.}\\
\parbox{.45\textwidth}{\hspace{0.2cm}{}\hspace{0.6cm}b) ${\lambda ^{{\cal A}}(k)} = \lambda$.}\\
\parbox{.45\textwidth}{\hspace{0.2cm}{{4.}}\hspace{0.4cm}a) Keep the ${\bm{\beta} ^{{\cal A}}}$ unchanged, optimize the ${\bm{\beta} ^{{\cal B}}}$ using \eqref{eq:objective_func_B}, i.e.,}\\
\parbox{.45\textwidth}{\hspace{0.8cm}{ } \hspace{0.2cm}${\bm{\beta} ^{{\cal B}}}{\rm{ = }}\mathop {{\rm{arg min}}}\limits_{\beta _{X_s}^{{\cal B}}} \left\{ {\sum\limits_{T_s} {D_{KL}^{{{\cal B}},T_s}}  - \lambda \left[ {\sum\limits_{X_s} {h(\beta _{X_s}^{{\cal B}}) - {L \mathord{\left/ {\vphantom {m 2}} \right. \kern-\nulldelimiterspace} 2}} } \right]} \right\}$.}\\
\parbox{.45\textwidth}{\hspace{0.2cm}{}\hspace{0.6cm}b) ${\lambda ^{{\cal B}}(k)} = \lambda$.}\\
\parbox{.45\textwidth}{\hspace{0.2cm}{{5.}}\hspace{0.4cm}\textbf{when} $k \ge 2$ }\\
\parbox{.45\textwidth}{\hspace{0.2cm}{}\hspace{1.0cm} $r^{{\cal A}} = {\lambda ^{{\cal A}}(k) / \lambda ^{{\cal A}}(k-1)}$,\hspace{0.2cm}$r^{{\cal B}} = {\lambda ^{{\cal B}}(k) / \lambda ^{{\cal B}}(k-1)}$.}\\
\parbox{.45\textwidth}{\hspace{0.2cm}{}\hspace{1.0cm} if ($r^{{\cal A}}>0.98$ \&\& $r^{{\cal B}}>0.98$), \hspace{0.2cm} then \textbf{return}.}\\
\parbox{.45\textwidth}{\hspace{0.2cm}{}\hspace{0.6cm}\textbf{end when}}\\

\parbox{.45\textwidth}{\hspace{0.2cm}{{6.}} \textbf{end} \textbf{for}} \\
\parbox{.45\textwidth}{\hspace{0.2cm}{{7.}} Compute the pixel embedding cost $d$ corresponding to ${\bm{\beta} ^{{\cal A}}}$ and ${\bm{\beta} ^{{\cal B}}}$:}\\
\parbox{.45\textwidth}{\hspace{0.2cm}\hspace{0.3cm} $d = \ln ({1 \mathord{\left/{\vphantom {1 {\beta  - 1}}} \right. \kern-\nulldelimiterspace} {\beta  - 2}}).$}\\
\parbox{.45\textwidth}{\hspace{0.2cm}{{8.}} {Embed $L$ bits into cover image $\bf{c}$ to obtain the stego image $\bf{s}$}.} \\
\bottomrule[1pt]
\end{tabular}
\end{table}
\section{Experimental Results and Analysis}
\subsection{Experiment setups}
All the experiments in this Section are carried out on image database BOSSbase ver1.01 \cite{bossbase} which contains 10,000 gray-scale images of size $512 \times 512 \times 8$ bits. All tested schemes are simulated at their corresponding payload-distortion bound, which execute the embedding modifications with the probabilities. The block size for variances and correlation coefficients estimation is $9\times9$, and the threshold for dynamical clique allocation is set as ${\beta _T} = 0.1$. Two state-of-the-art feature sets for spatial images, i.e., SRM \cite{SRM} and its selection-channel-aware version maxSRMd2 \cite{maxSRMd2} are used to evaluate the empirical security performance of the proposed GMRF based scheme and other competing methods. The Fisher Linear Discriminant ensemble \cite{ensemble} is also adopted in our experiments to train the binary classifier. Half of the cover and stego images will be used as the training set for the ensemble classifiers, and the remaining half will be used as test set to evaluate the trained classifier. And the security performance is quantified as the minimal total probability of error under equal priors achieved on the test set by ten times of randomly testing, denoted as ${\overline{P_E}}$.

\subsection{Performance comparison with other competing scheme}
We compare the proposed GMRF with the baseline of the state-of-the-art model based scheme MiPOD without smoothing on its Fisher information. Table \ref{tab:no-smoothing} show the security performance of the two schemes against SRM and maxSRMd2. It is observed that the GMRF consistently outperforms MiPOD by a clear margin for SRM across the tested payloads, indicating that the proposed GMRF can better preserve the statistical distribution of stego images after embedding and lead to less detectibility. For the selection-channel-aware maxSRMd2, however, our GMRF has slightly better performance than MiPOD for low payloads ($R \le 0.2{\rm{ bpp}}$), and tends to be inferior to MiPOD for medium and large payloads ($R \ge 0.3 {\rm{ bpp}}$), which is the range of insecurity (the corresponding $\overline{P_E} < 0.4$). This is most likely because the GMRF could better characterize the images than MiPOD, which can be acquired by maxSRMd2 through the change probabilities $\beta$ to more effectively detect the stego images.

\begin{table*}
\renewcommand\arraystretch{1.5}
\centering
\caption{Average testing error $\overline{P_E}$ versus payload in bits per pixel (bpp) for MiPOD and GMRF on BOSSbase ver1.01 using ensemble 1.0 classifier with two feature sets: SRM and maxSRMd2.}\label{tab:no-smoothing}
\begin{tabular}{cccccccc}
\toprule[1pt]
 \multirow{2}{*}{Feature} & \multirow{2}{*}{Algorithm} & \multicolumn{6}{c}{Payload(bpp)} \\
\cmidrule(lr){3-8}
  & &0.05 &0.1 &0.2 &0.3 &0.4 &0.5\\
\midrule[0.8pt]

\multirow{3}{*}{SRM} &MiPOD &$0.4511 \pm 0.0030$	&$0.4051 \pm 0.0041$	&$0.3274 \pm 0.0026$	&$0.2723 \pm 0.0041$	&$0.2218 \pm 0.0019$	&$0.1821 \pm 0.0023$\\
                     &GMRF  &$0.4558 \pm 0.0024$	&$0.4143 \pm 0.0036$	&$0.3428 \pm 0.0038$	&$0.2860 \pm 0.0043$	&$0.2350 \pm 0.0035$	&$ 0.1925 \pm 0.0028$\\
\midrule[0.8pt]

\multirow{3}{*}{maxSRMd2} &MiPOD &$0.4294 \pm 0.0037$	&$0.3772 \pm 0.0028$	&$0.3037 \pm 0.0028$	&$0.2498 \pm 0.0034$	&$0.2053 \pm 0.0028$	&$0.1683 \pm 0.0037$\\
                          &GMRF  &$0.4361 \pm 0.0031$	&$0.3830 \pm 0.0020$	&$0.3078 \pm 0.0036$	&$0.2467 \pm 0.0035$	&$0.1949 \pm 0.0041$	&$0.1595 \pm 0.0033$\\
\bottomrule[1pt]
\end{tabular}
\end{table*}

\subsection{The effect of smoothing operations on GMRF and MiPOD}
Nowadays, it is a common practice to boost the security performance of steganographic schemes by smoothing the embedding costs using a low-pass filter \cite{HiLL-J}. This can be explained from the perspective of  maximum entropy principle for discrete source in information theory that the smoothing operation on embedding costs tends to uniformize the change probabilities in local regions and thus increase the embedding entropy in highly textured regions. In addition, the smoothing operation can also spread the high costs of pixels into their neighborhood which would make the embedding more conservative in textured edges. In \cite{MiPOD}, the authors take a similar measure to improve the performance of MiPOD by smoothing the Fisher information, which is closely correlated with the embedding cost. While in the paper, for ensuring the fairness of the comparative experiment, we will choose to smooth the embedding cost of MiPOD directly, since smoothing the binary Fisher information in FIM for GMRF is inappropriate.

In our implementation, we adopt a low-pass filter with  support as the one used in MiPOD, i.e., a $7 \times 7$ average filter. Table \ref{tab:smoothing} show the performance comparison of HiLL, and low-pass filtered GMRF and MiPOD. It is observed that, for SRM, the proposed GMRF with low-pass filtered cost outperforms GMRF without low-pass filtering and MiPOD with low-pass filtered cost for the tested payloads as expected, but shows slightly inferior performance to HiLL. With the increase of payload, the performance gap between GMRF and HiLL becomes smaller and GMRF tends to exhibit comparable performance with HiLL for relatively large payload ($R$$\ge$$0.3{\rm bpp}$). For maxSRMd2, both GMRF and MiPOD show superior performance to HiLL, especially at small payload ($R$$\le$$0.2{\rm bpp}$), and the weakness of GMRF compared to MiPOD for relatively large payload ($R$$>$$0.3{\rm{ bpp}}$) is decreasing compared with the results in Table \ref{tab:no-smoothing}, which is due to the smoothing operation spreads out the change probabilities.

In short, the smoothing operations indeed boost the security performance of both GMRF and MiPOD against SRM and maxSRMd2, especially for maxSRMd2. On the other hand, the smoothing operation would attenuate the adaptability of GMRF, which is beneficial to the performance against maxSRMd2, and detrimental to the one against SRM. This explains why the performance improvement of GMRF (due to smoothing operation) for SRM is less than the one for maxSRMd2. Finally, although the proposed GMRF (with and without filtering) shows inferior performance to MiPOD for maxSRMd2 at relatively large payload, it exhibits superior security performance consistently to MiPOD for SRM. Considering the fact that, in practice, the precise knowledge of selection channels is generally unavailable to the steganalyzers, so the proposed GMRF is more preferable for practical applications compared with MiPOD.

\begin{table*}
\renewcommand\arraystretch{1.5}
\centering
\caption{Average testing error $\overline{P_E}$ versus payload in bits per pixel (bpp) for MiPOD with low-pass filtered costs, GMRF with low-pass filtered costs and HiLL on BOSSbase ver1.01 using ensemble 1.0 classifier with two feature sets: SRM and maxSRMd2.}\label{tab:smoothing}
\begin{tabular}{cccccccc}
\toprule[1pt]
 \multirow{2}{*}{Feature} & \multirow{2}{*}{Algorithm} & \multicolumn{6}{c}{Payload(bpp)} \\
\cmidrule(lr){3-8}
  & &0.05 &0.1 &0.2 &0.3 &0.4 &0.5\\
\midrule[0.8pt]

\multirow{2}{*}{SRM} &HiLL  &$0.4704 \pm 0.0024$	&$0.4330 \pm 0.0029$	&$0.3582 \pm 0.0026$	&$0.2975 \pm 0.0055$	&$0.2450 \pm 0.0025$	&$0.2018 \pm 0.0027$\\
                     &MiPOD &$0.4549 \pm 0.0034$	&$0.4139 \pm 0.0031$	&$0.3417 \pm 0.0031$	&$0.2811 \pm 0.0048$	&$0.2362 \pm 0.0039$	&$0.1910 \pm 0.0020$\\
                     &GMRF  &$0.4581 \pm 0.0027$	&$0.4210 \pm 0.0042$	&$0.3530 \pm 0.0031$	&$0.2948 \pm 0.0020$	&$0.2444 \pm 0.0033$	&$0.2015 \pm 0.0023$\\
\midrule[0.8pt]

\multirow{2}{*}{maxSRMd2} &HiLL  &$0.4237 \pm 0.0019$	&$0.3732 \pm 0.0045$	&$0.3094 \pm 0.0029$	&$0.2590 \pm 0.0029$	&$0.2169 \pm 0.0018$	&$0.1789 \pm 0.0033$\\
                          &MiPOD &$0.4416 \pm 0.0030$	&$0.3902 \pm 0.0028$	&$0.3231 \pm 0.0036$	&$0.2684 \pm 0.0026$	&$0.2212 \pm 0.0023$	&$0.1846 \pm 0.0029$\\
                          &GMRF  &$0.4445 \pm 0.0020$	&$0.3958 \pm 0.0032$	&$0.3240 \pm 0.0020$	&$0.2684 \pm 0.0035$	&$0.2201 \pm 0.0038$	&$0.1787 \pm 0.0024$\\

\bottomrule[1pt]
\end{tabular}
\end{table*}

\section{Conclusion}
At present, the prevailing methodology for adaptive image steganography is based on the framework of minimal distortion embedding, which includes the additive embedding cost for each cover element and the encoding method, typically syndrome-trellis codes (STCs), to minimize the sum of costs. Inspired by the recognition that the security performance of image steganography could be improved by taking advantages of the non-additive model, a Gaussian Markov Random Field (GMRF) with four-element cross neighborhood is proposed to capture the dependences among spatially adjacent pixels, and the problem of secure image steganography is formulated as the minimization of KL-divergence between cover and stego based on sound mathematical principle. The adoption of the proposed GMRF allows to effectively characterize the high-dimensional joint distribution of cover elements and the corresponding KL-divergence in terms of a series of low-dimensional clique structures. With the proposed GMRF, the cover image is tessellated into two disjoint subimages, which are conditionally independent. An alternating iterative optimization scheme is then developed to tackle the issue of efficient embedding while minimizing the total KL-divergence. Finally, the performance of the proposed GMRF is further boosted with smoothing operations on obtained costs. Experiments are carried out to demonstrate the superior performance of the proposed GMRF in terms of secure payload against steganalysis and show that the GMRF outperforms the prior arts, e.g., MiPOD, which is based on the multivariate independent Gaussian model, and has comparable performance with the state-of-the-art HiLL against SRM for tested payloads, where the selection-channel knowledges are unavailable to the steganalyzers and is more preferable for practical applications.

\appendices
\makeatletter
\@addtoreset{equation}{section}
\makeatother
\renewcommand{\theequation}{\Alph{section}-\arabic{equation}}

\section{KL-divergence and Fisher information matrix for a two-pixel clique}
\subsection*{1. KL-divergence}
Let  ${F_P({{\bf X}})}$ and $F_Q^{\bm{\beta }}({\bf Y})$ be the p.m.f of two-pixel cover clique ${\bf X} = [X_1, X_2]^T$ and stego clique ${\bf Y} = [Y_1, Y_2]^T$, respectively. And the associated change probabilities are ${\bm {\beta}} = [{\beta}_1, {\beta}_2]^T$, the KL-divergence $D(\bm {\beta})$ between ${\bf X}$ and ${\bf Y}$ is
\begin{equation}\label{eq:Apd_KL-1}
D({\bm{\beta }}) = {D_{KL}}\left( {{F_P}||F_Q^{\bm{\beta }}} \right) = \sum {{F_P}\log \displaystyle\frac{{{F_P}}}{{F_Q^{\bm{\beta }}}}}.
\end{equation}
By taking the Taylor expansion at ${\bm{\beta }} = [0, 0]^T$, we have
\begin{equation}\label{eq:Apd_KL-2}			
D({\bm{\beta }}) = D({\bm{0}}) + {{\bm{\beta }}^T} \cdot \nabla D({\bm{0}}) + \displaystyle\frac{1}{2} \cdot {{\bm{\beta }}^T} \cdot {\nabla ^2}D({\bm{0}}) \cdot {\bm{\beta }} + {\cal O}(\xi ).
\end{equation}
Note that $D({\bm{0}}) = 0$ and $\nabla D({\bm{0}}) = 0$, for small ${\bm{\beta }}$, $D({\bm{\beta }})$ can then be well approximated with its leading quadratic term
\begin{equation}\label{eq:Apd_KL-3}
D({\bm{\beta }}) \approx \displaystyle\frac{1}{2} \cdot {{\bm{\beta }}^T} \cdot {\nabla ^2}D({\bm{0}}) \cdot {\bm{\beta }},
\end{equation}
where ${\nabla ^2}D({\bm{0}})$ is the second-order partial derivatives defined as
\begin{equation}\label{eq:Apd_KL-4}
{\nabla ^2}D({\bm{0}}) = {\left. {\left[ {\begin{array}{*{20}{c}}
{\displaystyle\frac{{{\partial ^2}D({\bm{\beta }})}}{{\partial {\beta _1}^2}}}&{\displaystyle\frac{{{\partial ^2}D({\bm{\beta }})}}{{\partial {\beta _1}\partial {\beta _2}}}}\\
{\displaystyle\frac{{{\partial ^2}D({\bm{\beta }})}}{{\partial {\beta _2}\partial {\beta _1}}}}&{\displaystyle\frac{{{\partial ^2}D({\bm{\beta }})}}{{\partial {\beta _2}^2}}}
\end{array}} \right]} \right|_{{\bm{\beta }} = {\bm{0}}}}.
\end{equation}

We then proceed to the computation for each of the components of ${\nabla ^2}D({\bm{0}})$. Note that ${\left. {F_Q^{\bm{\beta }}({\bf{X}} = {\bf{x}})} \right|_{{\bf{x}} = [i,j]}} = {q_{i,j}}$, according to \eqref{eq:one_pixel_stego_probability}, it is easily verified that
\begin{equation}\label{eq:Apd_KL-5}
\begin{array}{l}
{\left. {\displaystyle\frac{{{\partial ^2}F_Q^{\bm{\beta }}({\bf{X}} = {\bf{x}})}}{{\partial {\beta _1}^2}}} \right|_{{\bf{x}} = [i,j]}} = \displaystyle\frac{{{\partial ^2}{q_{i,j}}}}{{\partial {\beta _1}^2}}{\rm{ = }}0\\
{\left. {\displaystyle\frac{{{\partial ^2}F_Q^{\bm{\beta }}({\bf{X}} = {\bf{x}})}}{{\partial {\beta _2}^2}}} \right|_{{\bf{x}} = [i,j]}} = \displaystyle\frac{{{\partial ^2}{q_{i,j}}}}{{\partial {\beta _2}^2}}{\rm{ = }}0
\end{array}.
\end{equation}
According to the definition in \eqref{eq:Apd_KL-1}, we have
\begin{equation}\label{eq:Apd_KL-6}
\begin{array}{l}
\displaystyle\frac{{{\partial ^2}D({\bm{\beta }})}}{{\partial {\beta _1}^2}} = \displaystyle\frac{{ - 1}}{{\ln 2}} \cdot \sum\limits_{\bf{x}} {{F_P}({\bf{X}}) \displaystyle\frac{\partial }{{\partial {\beta _1}}}\left( {\displaystyle\frac{1}{{F_Q^{\bm{\beta }}({\bf{X}})}} \cdot \displaystyle\frac{{\partial F_Q^{\bm{\beta }}({\bf{X}} )}}{{\partial {\beta _1}}}} \right)} \\
\hspace{0.5cm} = \displaystyle\frac{{ - 1}}{{\ln 2}} \sum\limits_{\bf{x}} {{F_P}({\bf{X}})\left( \begin{array}{l}
\displaystyle\frac{{ - 1}}{{{{\left( {F_Q^{\bm{\beta }}({\bf{X}})} \right)}^2}}} \cdot {\left( {\displaystyle\frac{{\partial F_Q^{\bm{\beta }}({\bf{X}})}}{{\partial {\beta _1}}}} \right)^2}\\
\hspace{0.5cm} + \hspace{0.3cm} \displaystyle\frac{1}{{F_Q^{\bm{\beta }}({\bf{X}})}}\displaystyle\frac{{{\partial ^2}F_Q^{\bm{\beta }}({\bf{X}} )}}{{\partial {\beta _1}^2}}
\end{array} \right)}
\end{array}.
\end{equation}
Note that ${\left. {{F_P}({\bf{X}}) = F_Q^{\bm{\beta }}({\bf{X}})} \right|_{{\bm{\beta }} = {\bf{0}}}}$, substituting \eqref{eq:Apd_KL-5} into \eqref{eq:Apd_KL-6} gives
\begin{equation}\label{eq:Apd_KL-7}
\begin{array}{l}
{\left. {\displaystyle\frac{{{\partial ^2}D({\bm{\beta }})}}{{\partial {\beta _1}^2}}} \right|_{_{{\bm{\beta }} = {\bf{0}}}}} = {\left. {\displaystyle\frac{1}{{\ln 2}} \cdot \sum\limits_{\bf{x}} {{F_P}({\bf{X}}){{\left( {\displaystyle\frac{1}{{F_Q^{\bm{\beta }}({\bf{X}})}} \cdot \displaystyle\frac{{\partial F_Q^{\bm{\beta }}({\bf{X}})}}{{\partial {\beta _1}}}} \right)}^2}} } \right|_{{\bm{\beta }} = {\bf{0}}}}\\
\hspace{0.5cm} = \displaystyle\frac{1}{{\ln 2}} \cdot {\left. {\sum\limits_{\bf{x}} {{F_P}({\bf{X}}){{\left( {\displaystyle\frac{{\partial \ln F_Q^{\bm{\beta }}({\bf{X}})}}{{\partial {\beta _1}}}} \right)}^2}} } \right|_{{\bm{\beta }} = {\bf{0}}}}\\
\hspace{0.5cm} = \displaystyle\frac{1}{{\ln 2}} \cdot E\left[ {{{\left. {{{\left( {\displaystyle\frac{{\partial \ln F_Q^{\bm{\beta }}({\bf{X}})}}{{\partial {\beta _1}}}} \right)}^2}} \right|}_{{\bm{\beta }} = {\bf{0}}}}} \right] = \displaystyle\frac{1}{{\ln 2}} \cdot {I_2}{({\bf{0}})_{1,1}}
\end{array},
\end{equation}
and similarly we can obtain
\begin{equation}\label{eq:Apd_KL-8}
{\left. {\displaystyle\frac{{{\partial ^2}D({\bm{\beta }})}}{{\partial {\beta _2}^2}}} \right|_{{\bm{\beta }} = {\bf{0}}}} = \displaystyle\frac{1}{{\ln 2}} \cdot E\left[ {{{\left. {{{\left( {\displaystyle\frac{{\partial \ln F_Q^{\bm{\beta }}({\bf{X}})}}{{\partial {\beta _2}}}} \right)}^2}} \right|}_{{\bm{\beta }} = {\bf{0}}}}} \right] = \displaystyle\frac{{{I_2}{{({\bf{0}})}_{{\rm{2}},{\rm{2}}}}}}{{\ln 2}}.
\end{equation}
As for  the two terms $\displaystyle\frac{{{\partial ^2}D({\bm{\beta }})}}{{\partial {\beta _1}\partial {\beta _2}}}$ and $\displaystyle\frac{{{\partial ^2}D({\bm{\beta }})}}{{\partial {\beta _2}\partial {\beta _1}}}$, we have
\begin{equation}\label{eq:Apd_KL-9}
\begin{array}{l}
\displaystyle\frac{{{\partial ^2}D({\bm{\beta }})}}{{\partial {\beta _1}\partial {\beta _2}}} = \displaystyle\frac{{{\partial ^2}D({\bm{\beta }})}}{{\partial {\beta _2}\partial {\beta _1}}}\\
 = \displaystyle\frac{{ - 1}}{{\ln 2}} \cdot \sum\limits_{\bf{x}} {{F_P}({\bf{X}}) \displaystyle\frac{\partial }{{\partial {\beta _2}}}\left( {\displaystyle\frac{1}{{F_Q^{\bm{\beta }}({\bf{X}})}} \cdot \displaystyle\frac{{\partial F_Q^{\bm{\beta }}({\bf{X}})}}{{\partial {\beta _1}}}} \right)} \\
 = \displaystyle\frac{{ - 1}}{{\ln 2}} \cdot \sum\limits_{\bf{x}} {{F_P}({\bf{X}}) \left( \begin{array}{l}
\displaystyle\frac{1}{{F_Q^{\bm{\beta }}({\bf{X}})}} \cdot \displaystyle\frac{{{\partial ^2}F_Q^{\bm{\beta }}({\bf{X}})}}{{\partial {\beta _1}\partial {\beta _2}}}\\
- \displaystyle\frac{1}{{{{\left(F_Q^{\bm{\beta }}({\bf{X}})\right)}^2}}} \cdot \displaystyle\frac{{\partial F_Q^{\bm{\beta }}({\bf{X}})}}{{\partial {\beta _2}}} \cdot \displaystyle\frac{{\partial F_Q^{\bm{\beta }}({\bf{X}})}}{{\partial {\beta _1}}}
\end{array} \right)}
\end{array},
\end{equation}
where
\begin{equation}\label{eq:Apd_KL-10}
\begin{array}{l}
{\left. {\displaystyle\frac{{{\partial ^2}F_Q^{\bm{\beta }}({\bf{X}} = {\bf{x}})}}{{\partial {\beta _1}\partial {\beta _2}}}} \right|_{{\bf{x}} = [i,j]}} = \displaystyle\frac{{{\partial ^2}{q_{i,j}}}}{{\partial {\beta _1}\partial {\beta _2}}}\\
 = 4{p_{i,j}} - 2({p_{i{\rm{ - 1}},j}} + {p_{i{\rm{ + 1}},j}} + {p_{i,j - 1}} + {p_{i,j{\rm{ + }}1}})\\
 + ({p_{i{\rm{ - 1}},j{\rm{ - 1}}}} + {p_{i{\rm{ - 1}},j{\rm{ + 1}}}} + {p_{i{\rm{ + 1}},j{\rm{ - 1}}}} + {p_{i{\rm{ + 1}},j{\rm{ + 1}}}})
\end{array}.
\end{equation}
Similarly, for symmetric embedding, we can verify that \eqref{eq:Apd_KL-10} is equal to $0$. Therefore
\begin{equation}\label{eq:Apd_KL-11}
\begin{array}{l}
{\left. {\displaystyle\frac{{{\partial ^2}D({\bm{\beta }})}}{{\partial {\beta _1}\partial {\beta _2}}}} \right|_{\bm{\beta }  = {\bf{0}}}} = {\left. {\displaystyle\frac{{{\partial ^2}D({\bm{\beta }})}}{{\partial {\beta _2}\partial {\beta _1}}}} \right|_{\bm{\beta }  = {\bf{0}}}}\\
 = \displaystyle\frac{1}{{\ln 2}} \cdot E\left[ {{{\left. {\left( {\displaystyle\frac{{\partial \ln F_Q^{\bm{\beta }}({\bf{X}})}}{{\partial {\beta _1}}} \cdot \displaystyle\frac{{\partial \ln F_Q^{\bm{\beta }}({\bf{X}})}}{{\partial {\beta _2}}}} \right)} \right|}_{\bm{\beta }  = {\bf{0}}}}} \right]\\
 = \displaystyle\frac{1}{{\ln 2}} \cdot {I_2}{({\bf{0}})_{1,{\rm{2}}}} = \displaystyle\frac{1}{{\ln 2}} \cdot {I_2}{({\bf{0}})_{{\rm{2}},1}}
\end{array}.
\end{equation}
The ${I_2}{({\bf{0}})_{1,{\rm{1}}}}$, ${I_2}{({\bf{0}})_{1,{\rm{2}}}}$, ${I_2}{({\bf{0}})_{2,{\rm{1}}}}$ and ${I_2}{({\bf{0}})_{2,{\rm{2}}}}$ in \eqref{eq:Apd_KL-7}, \eqref{eq:Apd_KL-8} and \eqref{eq:Apd_KL-11} are the components for $2 \times 2$ binary Fisher Information Matrix (FIM) ${{\bf{I}}_2}({\bf{0}})$ with respect to $\bf{X}$.

\subsection*{2. Fisher information}
Recognize that the KL-divergence for a two-pixel clique $\bf{X}$ can be represented in terms of its FIM, we then further seek to derive ${{\bf{I}}_2}({\bf{0}})$ for Gaussian distributed $\bf{X}$. Without loss of generality, we only give the derivation of ${I_2}{({\bf{0}})_{1,{\rm{2}}}}$, other components of FIM can be obtained in the same way. By definition,
\begin{equation}\label{eq:Apd_FI-1}
{I_2}{({\bf{0}})_{1,2}} = E\left[ {{{\left. {\left( {\displaystyle\frac{{\partial \ln F_Q^{\bm{\beta }}({\bf{X}})}}{{\partial {\beta _1}}} \cdot \displaystyle\frac{{\partial \ln F_Q^{\bm{\beta }}({\bf{X}})}}{{\partial {\beta _2}}}} \right)} \right|}_{{\bm{\beta }} = {\bf{0}}}}} \right].
\end{equation}
By substituting \eqref{eq:one_pixel_stego_probability} into \eqref{eq:Apd_FI-1} and let $\bm{\beta}  \to {\bf{0}}$, we can come to that
\begin{equation}\label{eq:Apd_FI-2}
\begin{array}{l}
{\left. {\displaystyle\frac{{\partial \ln F_Q^{\bm{\beta }}({\bf{X}=\bf{x}})}}{{\partial {\beta _1}}}} \right|_{{\bm{\beta }} = {\bf{0}},{\bf {x}}=[i,j]}} = \displaystyle\frac{{\left[ {{p_{i - 1,j}} + {p_{i + 1,j}} - 2{p_{i,j}}} \right]}}{{{q_{i,j}}}}\\
{\left. {\displaystyle\frac{{\partial \ln F_Q^{\bm{\beta }}({\bf{X}=\bf{x}})}}{{\partial {\beta _2}}}} \right|_{{\bm{\beta }} = {\bf{0}},{\bf {x}}=[i,j]}} = \displaystyle\frac{{\left[ {{p_{i,j - 1}} + {p_{i,j + 1}} - 2{p_{i,j}}} \right]}}{{{q_{i,j}}}}
\end{array}.
\end{equation}

Denote $\Omega _{i,j}^1 = ({p_{i - 1,j}} + {p_{i + 1,j}} - 2{p_{i,j}})$ and ${\Omega _{i,j}^2} = ({p_{i,j - 1}} + {p_{i,j + 1}} - 2{p_{i,j}})$, we then try to determine \eqref{eq:Apd_FI-2} by taking advantages of the Gaussian distribution $f({\bf X})$ for clique ${\bf X}$. According to \eqref{eq:MVT}, $p_{i,j}$ for clique ${\bf {x}} = [x_1, x_2]$ can be formulated as
\begin{equation}\label{eq:Apd_FI-3}
{p_{i,j}} = \left. {{F_\Delta }({x_1},{x_2})} \right|_{x_1=i\Delta,x_2=j\Delta} = {\Delta ^2} \cdot {f}(i'\Delta ,j'\Delta ),
\end{equation}
where ${f}({x_1},{x_2})$ is the bivariate Gaussian p.d.f. (see (5)), $i' \in (i-0.5,i+0.5)$ and $j' \in (j-0.5,j+0.5)$.  For fine quantization step $\Delta$, ${f}(i'\Delta ,j'\Delta ) \approx {f}(i\Delta ,j\Delta )$ , thus we can obtain ${p_{i \pm 1,j}}$ and ${p_{i,j \pm 1}}$ through the Taylor expansion of ${\Delta ^2} \cdot {f}({x_{1}},{x_{2}})$ at ${x_1} = i\Delta $ and ${x_2} = j\Delta $ , i.e.,
\begin{equation}\label{eq:Apd_FI-4}
\begin{array}{l}
{p_{i \pm 1,j}} = {F_\Delta }((i \pm 1)\Delta ,j\Delta ) \approx {\Delta ^2} \cdot {f}((i \pm 1)\Delta ,j\Delta )\\
= {\Delta ^2} \cdot {f}(i\Delta ,j\Delta ) + {\Delta ^2} \cdot \sum\limits_{l = 1}^\infty  {\displaystyle\frac{{{{( \pm \Delta )}^l}}}{{l!}} \cdot {{\left. {\displaystyle\frac{{{\partial ^l}{f}({x_1},{x_2})}}{{\partial {{x_1}^l}}}} \right|}_{{x_1} = i\Delta}}}
\end{array},
\end{equation}
\begin{equation}\label{eq:Apd_FI-5}
\begin{array}{l}
{p_{i,j \pm 1}} = {F_\Delta }(i\Delta ,(j \pm 1)\Delta ) \approx {\Delta ^2} \cdot {f}(i\Delta ,(j \pm 1))\\
= {\Delta ^2} \cdot {f}(i\Delta ,j\Delta ) + {\Delta ^2} \cdot {\sum\limits_{l = 1}^\infty  {\displaystyle\frac{{{{( \pm \Delta )}^l}}}{{l!}} \cdot \left. {\displaystyle\frac{{{\partial ^l}{f}({x_1},{x_2})}}{{\partial {{x_2}^l}}}} \right|} _{{x_2} = j\Delta }}
\end{array}.
\end{equation}
Based on these, $\Omega _{i,j}^1$ can then be formulated as
\begin{equation}\label{eq:Apd_FI-6}
\begin{array}{l}
\Omega _{i,j}^1 \approx 2{\Delta ^2} \cdot {f}(i\Delta ,j\Delta ) \\
+ {\Delta ^2} \cdot {\sum\limits_{l = 1}^3 {\displaystyle\frac{{{{( + \Delta )}^l}}}{{l!}} \cdot \left. {\displaystyle\frac{{{\partial ^l}{f}({x_1},{x_2})}}{{\partial {{x_1}^l}}}} \right|} _{{x_1} = i\Delta, {x_2} = j\Delta}}\\
 + {\Delta ^2} \cdot \sum\limits_{l = 1}^3 {\displaystyle\frac{{{{( - \Delta )}^l}}}{{l!}} \cdot {{\left. {\displaystyle\frac{{{\partial ^l}{f}({x_1},{x_2})}}{{\partial {{x_2}^l}}}} \right|}_{{x_1} = i\Delta, {x_2} = j\Delta }}}\\
 - 2{\Delta ^2} \cdot {f}({x_{1}},{x_{2}}) + {{\cal O}}({\Delta ^4},{\Delta ^4})
\end{array}.
\end{equation}
Therefore, \eqref{eq:Apd_FI-6} can be simplified as
\begin{equation}\label{eq:Apd_FI-7}
\Omega _{i,j}^1 \approx {\Delta ^4} \cdot {\left. {\displaystyle\frac{{\partial {}^2{f}({x_1},{x_2})}}{{\partial {{x_1}^2}}}} \right|_{{x_1} = i\Delta ,{x_2} = j\Delta }},
\end{equation}
and similarly, we have
\begin{equation}\label{eq:Apd_FI-8}
\Omega _{i,j}^2 \approx {\Delta ^4} \cdot {\left. {\displaystyle\frac{{\partial {}^2{f}({x_1},{x_2})}}{{\partial {{x_2}^2}}}} \right|_{{x_1} = i\Delta ,{x_2} = j\Delta }}.
\end{equation}
Substitute \eqref{eq:Apd_FI-2} into \eqref{eq:Apd_FI-1} and note that $q_{i,j}=p_{i,j}$ when ${\bm {\beta}} = \bf {0}$, we can finally obtain
\begin{equation}\label{eq:Apd_FI-9}
\begin{array}{l}
{I_2}{({\bf{0}})_{1,2}} = {I_2}{({\bf{0}})_{2,1}} = E\left[ {{{\left. {\displaystyle\frac{{\Omega _{i,j}^1 \cdot \Omega _{i,j}^2}}{{{{q_{i,j}}^2}}}} \right|}_{{\bm{\beta }} = {\bf{0}}}}} \right]\\
{\rm{ = }}{\left. {\sum\limits_i {\sum\limits_j {\displaystyle\frac{{\Omega _{i,j}^1 \cdot \Omega _{i,j}^2}}{{{p_{i,j}}}}} } } \right|_{{\bm{\beta }} = {\bf{0}}}}
\end{array}.
\end{equation}
Similarly, the other two components of ${\bf I}_2(\bf 0)$ are
\begin{equation}\label{eq:Apd_FI-10}
{\left. {{I_2}{{({\bf{0}})}_{1,1}} = \sum\limits_i {\sum\limits_j {\displaystyle\frac{{{{(\Omega _{i,j}^1)}^2}}}{{{p_{i,j}}}}} } } \right|_{{\bm{\beta }} = {\bf{0}}}},
\end{equation}
\begin{equation}\label{eq:Apd_FI-11}
{\left. {{I_2}{{({\bf{0}})}_{2,2}} = \sum\limits_i {\sum\limits_j {\displaystyle\frac{{{{(\Omega _{i,j}^2)}^2}}}{{{p_{i,j}}}}} } } \right|_{{\bm{\beta }} = {\bf{0}}}}.
\end{equation}
{{\bf{\emph{Proof for} \eqref{eq:Apd_KL-10}}:}
As an supplement, refer to \eqref{eq:Apd_FI-3}, \eqref{eq:Apd_FI-4} and \eqref{eq:Apd_FI-5}, we can further obtain ${p_{i \pm 1,j \pm 1}} = {F_\Delta }((i \pm 1)\Delta ,(j \pm 1)\Delta )$ using Taylor expansion of ${\Delta ^2} \cdot {f}({x_{1}},{x_{2}})$ at ${x_1} = i\Delta $ and ${x_2} = j\Delta $,  simultaneously. For simplicity, we expand it to the second order, and take ${p_{i{\rm{ + }}1,j + 1}}={F_\Delta }((i + 1)\Delta ,(j + 1)\Delta )$ for example, i.e.,
\begin{equation}\label{eq:Apd_FI-12}
\begin{array}{l}
{p_{i{\rm{ + }}1,j{\rm{ + }}1}} \approx {\Delta ^2} \cdot f((i + 1)\Delta ,(j + 1)\Delta )\\
 = {\Delta ^{\rm{2}}} \cdot \Delta  \cdot {\left. {\left( {\displaystyle\frac{{\partial f({x_1},{x_2})}}{{\partial {x_1}}}{\rm{ + }}\displaystyle\frac{{\partial f({x_1},{x_2})}}{{\partial {x_2}}}} \right)} \right|_{{x_1} = i\Delta ,{x_2} = j\Delta }}\\
 + \displaystyle\frac{{{\Delta ^{\rm{2}}} \cdot {\Delta ^{\rm{2}}}}}{2} \cdot {\left. {\left( {\displaystyle\frac{{{\partial ^{\rm{2}}}f({x_1},{x_2})}}{{\partial {{({x_1})}^2}}}{\rm{ + }}\displaystyle\frac{{{\partial ^{\rm{2}}}f({x_1},{x_2})}}{{\partial {{({x_{\rm{2}}})}^2}}}} \right)} \right|_{{x_1} = i\Delta ,{x_2} = j\Delta }}\\
 + {\Delta ^{\rm{2}}} \cdot \Delta  \cdot \Delta  \cdot {\left. {\displaystyle\frac{{{\partial ^{\rm{2}}}f({x_1},{x_2})}}{{\partial {x_1}\partial {x_2}}}} \right|_{{x_1} = i\Delta ,{x_2} = j\Delta }}{\rm{ + }}{\Delta ^2} \cdot f(i\Delta ,j\Delta )
\end{array}.
\end{equation}
Similarly, we can get ${p_{i - 1,j - 1}},{p_{i - 1,j{\rm{ + }}1}},{p_{i{\rm{ + }}1,j - 1}}$, then we have
\begin{equation}\label{eq:Apd_FI-13}
\begin{array}{l}
{p_{i - 1,j - 1}} + {p_{i - 1,j{\rm{ + }}1}} + {p_{i{\rm{ + }}1,j - 1}} + {p_{i{\rm{ + }}1,j{\rm{ + }}1}}\\
{\rm{ = 4}}{\Delta ^2} \cdot f(i\Delta ,j\Delta )\\
 + {\rm{2}}{\Delta ^{\rm{4}}} \cdot {\left. {\left( {\displaystyle\frac{{{\partial ^{\rm{2}}}f({x_1},{x_2})}}{{\partial {{({x_1})}^2}}}{\rm{ + }}\displaystyle\frac{{{\partial ^{\rm{2}}}f({x_1},{x_2})}}{{\partial {{({x_{\rm{2}}})}^2}}}} \right)} \right|_{{x_1} = i\Delta ,{x_2} = j\Delta }}
\end{array}.
\end{equation}
Next, according to \eqref{eq:Apd_FI-4} and \eqref{eq:Apd_FI-5}, we have
\begin{equation}\label{eq:Apd_FI-14}
\begin{array}{l}
{p_{i - 1,j}} + {p_{i{\rm{ + }}1,j}} + {p_{i,j - 1}} + {p_{i,j{\rm{ + }}1}}\\
{\rm{ = 4}}{\Delta ^2} \cdot f(i\Delta ,j\Delta )\\
 + {\Delta ^{\rm{4}}} \cdot {\left. {\left( {\displaystyle\frac{{{\partial ^2}f({x_1},{x_2})}}{{\partial {{({x_1})}^2}}} + \displaystyle\frac{{{\partial ^2}f({x_1},{x_2})}}{{\partial {{({x_2})}^2}}}} \right)} \right|_{{x_1} = i\Delta ,{x_2} = j\Delta }}
\end{array}.
\end{equation}
Finally, we substitute \eqref{eq:Apd_FI-13} and \eqref{eq:Apd_FI-14} in \eqref{eq:Apd_KL-10}, then we can obtain that the result is $0$.}

\section{KL-divergence for a 4-ary clique tree}
For the GMRF model with four-element cross neighborhood as shown in Fig. \ref{fig:GMRF_model}, the four neighboring pixels $\left\{ {{X_1},{X_2},{X_3},{X_4}} \right\}$ of ${X_s}$ are mutually independent when ${X_s}$ is given, which constitutes a {\textbf{4-ary clique tree}} ${T_s}$ centered at ${X_s}$. And the joint p.m.f. for ${T_s}$ is determined as
\begin{equation}\label{eq:Apd_KL_TREE-1}
\begin{array}{l}
{F_{{T_s}}}({{\bf{X}}_{{T_s}}}) = F({{\bf{X}}_{{T_s} - \{ {X_s}\} }}|{X_s}) \cdot F({X_s})\\
\hspace{0.2cm} = F({X_s}) \cdot \prod\limits_{t = 1}^4 {F({X_t}|{X_s})}  = \displaystyle\frac{{\prod\limits_{c \in {\cal C}} {F({\bf{X}_c})} }}{{{{\left(F({X_s})\right)}^3}}}
\end{array},
\end{equation}
where ${{\cal C}} = \left\{ {\{ {X_s},{X_t}\} |t = 1, \cdots ,4} \right\}$, $F({{\bf{X}}_c})$ and $F({X_s})$ are the p.m.f. for clique $c = \{ {X_s},{X_t}\}$ and pixel ${X_s}$, respectively. For cover clique tree  , the embedding modifies it to the stego one with change probabilities ${\bm{\beta }_{{T_s}}} = \left\{ {{\beta _{X_s}},{\beta _{X_1}},{\beta _{X_2}},{\beta _{X_3}},{\beta _{X_4}}} \right\}$, by applying \eqref{eq:Apd_KL_TREE-1}, the KL-divergence between cover and stego clique tree can be written as
\begin{equation}\label{eq:Apd_KL_TREE-2}
\begin{array}{l}
D_{KL}^{{T_s}}({F_P}||F_Q^{{{\bm{\beta }}_{{T_s}}}}{\rm{) = }}\sum\limits_{{{\bf{x}}_{{T_s}}}} {{F_P}({{\bf{X}}_{{T_s}}}) \log \displaystyle\frac{{{F_P}({{\bf{X}}_{{T_s}}})}}{{F_Q^{{{\bm{\beta }}_{{T_s}}}}({{\bf{X}}_{{T_s}}})}}} \\
 = \sum\limits_{{{\bf{x}}_{{T_s}}}} {{F_P}({{\bf{X}}_{{T_s}}}) \left[ {\log \displaystyle\frac{{\prod\limits_{c \in {\cal C}} {{F_P}({{\bf{X}}_c})} }}{{\prod\limits_{c \in {\cal C}} {F_Q^{{{\bm{\beta }}_c}}{\rm{(}}{{\bf{X}}_c}{\rm{)}}} }} - 3\cdot\log \displaystyle\frac{{{F_P}({X_s})}}{{F_Q^{{\beta _{X_s}}}{\rm{(}}{X_s}{\rm{)}}}}} \right]} \\
 = \overbrace {\sum\limits_{{{\bf{x}}_{{T_s}}}} {{F_P}({{\bf{X}}_{{T_s}}}) \log \displaystyle\frac{{\prod\limits_{c \in {\cal C}} {{F_P}({{\bf{X}}_c})} }}{{\prod\limits_{c \in {\cal C}} {F_Q^{{{\bm{\beta }}_c}}{\rm{(}}{{\bf{X}}_c}{\rm{)}}} }}} }^{\left\langle 1 \right\rangle }\\
\hspace{0.4cm} -3 \cdot \overbrace { \sum\limits_{{{\bf{x}}_{{T_s}}}} {{F_P}({{\bf{X}}_{{T_s}}}) } \log \displaystyle\frac{{{F_P}({X_s})}}{{F_Q^{{\beta _{X_s}}}{\rm{(}}{X_s}{\rm{)}}}}}^{\left\langle 2 \right\rangle }
\end{array}.
\end{equation}
For the first term ${\left\langle 1 \right\rangle }$ in \eqref{eq:Apd_KL_TREE-2}, we have
\begin{equation}\label{eq:Apd_KL_TREE-3}
\begin{array}{l}
\left\langle 1 \right\rangle  = {\rm{ }}\sum\limits_{{{\bf{x}}_{{T_s}}}} {\left( {\displaystyle\frac{{\prod\limits_{c \in {\cal C}} {{F_P}({{\bf{X}}_c})} }}{{{{\left( {{F_P}({X_s})} \right)}^3}}} \cdot \log \displaystyle\frac{{\prod\limits_{c \in {\cal C}} {{F_P}({{\bf{X}}_c})} }}{{\prod\limits_{c \in {\cal C}} {F_Q^{{{\bm{\beta }}_c}}{\rm{(}}{{\bf{X}}_c}{\rm{)}}} }}} \right)} \\
 = \sum\limits_{{{\bf{X}}_{{T_s}}}} {\left( {\displaystyle\frac{1}{{{{\left( {{F_P}({X_s})} \right)}^3}}} \cdot \prod\limits_{c \in {\cal C}} {{F_P}({{\bf{X}}_c})}  \cdot \sum\limits_{c \in {\cal C}} {\log \displaystyle\frac{{{F_P}({{\bf{X}}_c})}}{{F_Q^{{{\bm{\beta }}_c}}{\rm{(}}{{\bf{X}}_c}{\rm{)}}}}} } \right)}
\end{array}.
\end{equation}
For the summation $\sum\limits_{c \in {\cal C}} {\log \displaystyle\frac{{{F_P}({{\bf{X}}_c})}}{{F_Q^{{{\bm{\beta }}_c}}{\rm{(}}{{\bf{X}}_c}{\rm{)}}}}}$ in \eqref{eq:Apd_KL_TREE-3}, we take term associated with clique $c = \{ {X_s},{X_1}\}$ for example, other terms can be obtained similarly. We have
\begin{equation}\label{eq:Apd_KL_TREE-4}
\begin{array}{l}
\sum\limits_{{{\bf{x}}_{{T_s}}}} {\left( {\displaystyle\frac{1}{{{{\left( {{F_P}({X_s})} \right)}^3}}} \cdot \prod\limits_{c \in {\cal C}} {{F_P}({{\bf{X}}_c})}  \cdot \log \displaystyle\frac{{{F_P}({X_s},{X_1})}}{{F_Q^{{{\bm {\beta}} _{X_s,X_1}}}({X_s},{X_1})}}} \right)} \\
 = \sum\limits_{{x_s}} {\sum\limits_{{x_1}} {\displaystyle\left( {{F_P}({X_s},{X_1})\log \displaystyle\frac{{{F_P}({X_s},{X_1})}}{{F_Q^{{{\bm \beta} _{X_s,X_1}}}({X_s},{X_1})}} \cdot \Psi } \right)} }
\end{array},
\end{equation}
where $\Psi  = \left( {\sum\limits_{{x_2}} {\sum\limits_{{x_3}} {\sum\limits_{{x_4}} {\displaystyle\frac{{{F_P}({X_s},{X_2}){F_P}({X_s},{X_3}){F_P}({X_s},{X_4})}}{{{{\left( {{F_P}({X_s})} \right)}^3}}}} } } } \right)$.
It is easily verified that $\sum\limits_{{X_t}} {{F_P}({X_s},{X_t}) = {F_P}({X_s})}$, then $\Psi=1$, thus \eqref{eq:Apd_KL_TREE-4} can be simplified as the KL-divergence of clique $c = \{ {X_s},{X_1}\}$, that is
\begin{equation}\label{eq:Apd_KL_TREE-5}
{D_{KL}}({F_P}|{\rm{|F}}_Q^{{{\bm{\beta }}_{{X_s},{X_1}}}}) = \displaystyle\frac{{{\bm{\beta }}_{{X_s},{X_1}}^T{{\bf{I}}_{2,\{ {X_s},{X_1}\} }}({\bf{0}}){{\bm{\beta }}_{{X_s},{X_1}}}}}{{2\ln 2}}.
\end{equation}
On the other hand, the term $\left\langle 2 \right\rangle $ in \eqref{eq:Apd_KL_TREE-2} can also be simplified as the KL-divergence of pixel ${X_s}$ according to \cite{MiPOD}, i.e.,
\begin{equation}\label{eq:Apd_KL_TREE-6}
{D_{KL}}({F_P}||F_Q^{{\beta _{{X_{\rm{s}}}}}}) = {{{I_{1,{X_s}}}(0){{({\beta _{{X_{\rm{s}}}}})}^2}} \mathord{\left/
 {\vphantom {{{I_{1,{X_s}}}(0){{({\beta _{{X_{\rm{s}}}}})}^2}} {2\ln 2}}} \right.
 \kern-\nulldelimiterspace} {2\ln 2}}.
\end{equation}
Finally, the KL-divergence for 4-ary clique tree $T_s$ can then be written as
\begin{equation}\label{eq:Apd_KL_TREE-7}
\begin{array}{l}
D_{KL}^{{T_s}}({F_P}({{\bf{x}}_{{T_s}}})||F_Q^{{{\bm{\beta }}_{{T_s}}}}{\rm{(}}{{\bf{X}}_{{T_s}}}{\rm{))}}\\
{\rm{ = }}\sum\limits_{c \in {\cal C}} {{D_{KL}}} ({F_P}({{\bf{X}}_c})||F_Q^{{{\bm{\beta }}_{{T_s}}}}({{\bf{X}}_c})) - 3{D_{KL}}({F_P}({X_s})|{\rm{|F}}_Q^{{\beta _{{X_s}}}}({X_s}))\\
 = {{\left( {\sum\limits_{c \in {\cal C}} {{\bm{\beta }}_c^T{{\bf{I}}_{2,c}}({\bf{0}}){{\bm{\beta }}_c}}  - 3 \cdot {I_{1,{X_s}}}(0){{({\beta _{{X_s}}})}^2}} \right)} \mathord{\left/
 {\vphantom {{\left( {\sum\limits_{c \in {\cal C}} {{\bm{\beta }}_c^T{{\bf{I}}_{2,c}}({\bf{0}}){{\bm{\beta }}_c}}  - 3 \cdot {I_{1,{X_s}}}(0){{({\beta _{{X_s}}})}^2}} \right)} {2\ln 2}}} \right.
 \kern-\nulldelimiterspace} {2\ln 2}}.
\end{array}
\end{equation}
For practical steganography with dynamical clique allocation (see Section III-C), the KL-divergence in \eqref{eq:Apd_KL_TREE-7} should be rewritten as
\begin{equation}\label{eq:Apd_KL_TREE-8}
D_{KL}^{{T_s}} = \displaystyle\frac{{\sum\limits_{c \in {{\cal C}}} {([{\theta _c}]{\bm{\beta }}_c^T{{\bf{I}}_{2,c}}({\bf{0}}){{\bm{\beta }}_c}} ) - ((\sum\limits_{c \in {{\cal C}}} {[{\theta _c}]} ) - 1){I_{1,{X_s}}}(0){{\beta _{X_s}}^2}}}{{2\ln 2}},
\end{equation}
where $[{\theta _c}] = 1$ when clique $c$ is included in $T_s$ and zero otherwise.

\section{Minimization of total KL-divergence between cover and Stego with payload constraint}
The minimum distortion embedding with the proposed GMRF model can be formulated as the total KL-divergence minimization of sub-images ${\cal A}$ and ${\cal B}$ with payload constraints. We take the optimization on sub-image ${\cal A}$ for example, by applying the method of Lagrange multipliers, we have
\begin{equation}\label{eq:Apd_Opti_obj-1}
\mathop {min}\limits_{\beta _{{X_s}}^{{\cal A}}} \hspace{0.3cm}  \sum\limits_{{T_s}} {D_{KL}^{{{\cal A}},{T_s}}}  - \lambda \left[ {\sum\limits_{{X_s}} {h(\beta _{{X_s}}^{{\cal A}}) - {L \mathord{\left/
 {\vphantom {L 2}} \right.
 \kern-\nulldelimiterspace} 2}} } \right],
\end{equation}
where $D_{KL}^{{{\cal A}},{T_s}}$ is the KL-divergence for $s^{th}$ 4-ary clique tree $T_s$ centered at pixel $X_s$ in ${\cal A}$, ${\beta _{{X_s}}^{{\cal A}}}$ is the change probability for $X_s$ which we are going to optimize, and $L/2$ is the payload assigned to ${\cal A}$. Differentiating the objective function with respect to ${\beta _{{X_s}}^{{\cal A}}}$ gives
\begin{equation}\label{eq:Apd_Opti_obj-2}
\begin{array}{l}
\sum\limits_{{T_s}} {\left( {\displaystyle\frac{{\partial D_{KL}^{{{\cal A}},{T_s}}}}{{\partial \beta _{X_s}^{{\cal A}}}} - \lambda  \cdot \displaystyle\frac{{\partial h(\beta _{X_s}^{{\cal A}})}}{{\partial \beta _{X_s}^{{\cal A}}}}} \right) = 0} \\
\hspace{2.3cm} \Downarrow \\
\forall {T_s}:\overbrace {\displaystyle\frac{{\partial D_{KL}^{{{\cal A}},{T_s}}}}{{\partial \beta _{X_s}^{{\cal A}}}}}^{\left\langle 1 \right\rangle } - \lambda  \cdot \overbrace {\displaystyle\frac{{\partial h(\beta _{X_s}^{{\cal A}})}}{{\partial \beta _{X_s}^{{\cal A}}}}}^{\left\langle 2 \right\rangle } = 0
\end{array}.
\end{equation}
For term $\left\langle 1 \right\rangle$ in \eqref{eq:Apd_Opti_obj-2}, substituting \eqref{eq:Apd_KL_TREE-8} into $\left\langle 1 \right\rangle$ gives
\begin{equation}\label{eq:Apd_Opti_obj-3}
\begin{array}{l}
\displaystyle\frac{{\partial D_{KL}^{{{\cal A}},{T_s}}}}{{\partial \beta _{{X_s}}^{{\cal A}}}} = \displaystyle\frac{{\partial \left( \begin{array}{l}
\sum\limits_{c \in {{\cal C}}} {[\theta _c^{{T_s}}]{\bm{\beta }}_c^T{\bf{I}}_{2,c}^{{T_s}}({\bf{0}}){{\bm{\beta }}_c}} \\
\hspace{0.5cm} - \hspace{0.3cm} ((\sum\limits_{c \in {\cal C}} {[\theta _c^{{T_s}}]} ) - 1) \cdot I_{1,X_s}^{{T_s}}(0){\beta _{{X_s}}^{{\cal A}}}^2
\end{array} \right)}}{{2\ln 2 \cdot \partial \beta _{{X_s}}^{{\cal A}}}}\\
 = \displaystyle\frac{1}{{\ln 2}} \cdot \beta _{{X_s}}^{{\cal A}} \cdot \left( \begin{array}{l}
(\sum\limits_{c \in {{\cal C}}} {[\theta _c^{{T_s}}]I_{2,c}^{{T_s}}{{({\bf{0}})}_{2,2}}} )\\
\hspace{0.5cm} - \hspace{0.3cm}  ((\sum\limits_{c \in {{\cal C}}} {[\theta _c^{{T_s}}]} ) - 1) \cdot I_{1,X_s}^{{T_s}}(0)
\end{array} \right)\\
 + \displaystyle\frac{1}{{\ln 2}} \cdot \sum\limits_{c \in {{\cal C}}} {[\theta _c^{{T_s}}]I_{2,c}^{{T_s}}{{({\bf{0}})}_{1,2}}\beta _{{X_{s\_c}}}^{{{\cal B}},{T_s}}}
\end{array},
\end{equation}
where $\beta _{X_{s\_c}}^{{{\cal B}},{T_s}}$ is the change probability for pixel $X_{s\_c}$ corresponding to the clique $c = \{ {X_s},{X_{s\_c}}\}$ of 4-ary clique tree $T_s$, which is located in sub-image ${\cal B}$ and is fixed in the optimization process, ${\bf{I}}_{2,c}^{{T_s}}({\bf{0}})$ and $I_{1,X_s}^{{T_s}}(0)$ are the binary FIM for clique $c$ and FI associated with the $s^{th}$ pixel in ${\cal A}$, respectively. As for the term $\left\langle 2 \right\rangle$ in \eqref{eq:Apd_Opti_obj-2},we have
\begin{equation}\label{eq:Apd_Opti_obj-4}
\begin{array}{l}
\displaystyle\frac{{\partial h(\beta _{{X_s}}^{{\cal A}})}}{{\partial \beta _{{X_s}}^{{\cal A}}}} =  - \displaystyle\frac{{\partial (2\beta _{{X_s}}^{{\cal A}}\ln \beta _{{X_s}}^{{\cal A}} + (1 - 2\beta _{{X_s}}^{{\cal A}})\ln (1 - 2\beta _{{X_s}}^{{\cal A}}))}}{{\ln 2 \cdot \partial \beta _{{X_s}}^{{\cal A}}}}\\
 = \displaystyle\frac{2}{{\ln 2}} \cdot \ln \displaystyle\frac{{1 - 2\beta _{{X_s}}^{{\cal A}}}}{{\beta _{{X_s}}^{{\cal A}}}}.
\end{array}
\end{equation}
Substituting \eqref{eq:Apd_Opti_obj-3} and \eqref{eq:Apd_Opti_obj-4} into \eqref{eq:Apd_Opti_obj-2}, the we can finally obtain
\begin{equation}\label{eq:Apd_Opti_obj-5}
{\Gamma _{{X_s}}}\beta _{{X_s}}^{{\cal A}} + {\Lambda _{{X_s}}} - 2\lambda \ln \displaystyle\frac{{1 - 2\beta _{{X_s}}^{{\cal A}}}}{{\beta _{{X_s}}^{{\cal A}}}} = 0,
\end{equation}
where ${\Gamma _{{X_s}}} = \sum\limits_{c \in {{\cal C}}} {[\theta _c^{{T_s}}]I_{2,c}^{{T_s}}{{({\bf{0}})}_{2,2}} - (} (\sum\limits_{c \in {{\cal C}}} {[\theta _c^{{T_s}}]} ) - 1)I_{1,X_s}^{{T_s}}(0)$ and ${\Lambda _{{X_s}}} = \sum\limits_{c \in {{\cal C}}} {[\theta _c^{{T_s}}]I_{2,c}^{{T_s}}{{({\bf{0}})}_{1,2}}} \beta _{{X_{s\_c}}}^{{\cal B}}$. And \eqref{eq:Apd_Opti_obj-5} could be solved numerically for each $X_s$.

\ifCLASSOPTIONcaptionsoff
  \newpage
\fi

\bibliographystyle{IEEEtran}
\bibliography{mybibfile}

\end{document}